\documentclass[pre,showpacs,floatfix,onecolumn]{revtex4}

\usepackage{graphicx}
\usepackage{subfigure}
\usepackage{amsmath}
\usepackage{amssymb}
\usepackage{latexsym}
\usepackage{multirow}
\usepackage{color}
\usepackage[hidelinks]{hyperref}

\graphicspath{{./fig/}}

\newcommand{\etal}{{\sl et al. }}
\begin{document}

\title{The narrow escape problem in a circular domain with radial piecewise constant diffusivity}
\author{M. Mangeat}
\author{H. Rieger}

\affiliation{Center for Biophysics \& Department for Theoretical Physics, Saarland University, 66123 Saarbr\"ucken, Germany}

\begin{abstract}
The stochastic motion of particles in living cells is often spatially inhomogeneous with a higher effective diffusivity in a region close to the cell boundary due to active transport along actin filaments. As a first step to understand the consequence of the existence of two compartments with different diffusion constant for stochastic search problems we consider here a Brownian particle in a circular domain with different diffusion constants in the inner and the outer shell. We focus on the narrow escape problem and compute the mean first passage time (MFPT) for Brownian particles starting at some pre-defined position to find a small region on the outer reflecting boundary. For the annulus geometry we find that the MFPT can be minimized for a specific value of the width of the outer shell. In contrast for the two-shell geometry we show that the MFPT depends monotonously on all model parameters, in particular on the outer shell width. Moreover we find that the distance between the starting point and the narrow escape region which maximizes the MFPT depends discontinuously on the ratio between inner and outer diffusivity.
\end{abstract}

\pacs{}

\maketitle

\section{Introduction}

The narrow escape problem is ubiquitous in biology and chemistry problems~\cite{schuss_2007_narrow, bressloff_2013_stochastic, chou_2014_first, holcman_2014_narrow, iyer-biswas_2015_first} and consists in calculating the mean first passage time (MFPT) of a random walk or a Brownian motion from a starting point to a small absorbing window, the so-called escape or target region, on an otherwise reflecting boundary of a bounded domain.  Mathematically the problem reduces to solving a Poisson equation with mixed Neumann-Dirichlet boundary conditions~\cite{redner_2001_guide}. These equations can be solved numerically with a partial differential equation (PDE) solver for any complex geometry boundaries. Moreover, new numerical techniques allow the efficient simulation of the stochastic process using Kinetic Monte Carlo algorithms~\cite{schwarz_2013_efficient, schwarz_2016_optimality, schwarz_2016_numerical}.

Analytically, the narrow escape problem has been widely studied in the last decades~\cite{holcman_2004_escape, kolokolnikov_2005_optimizing, singer_2006_narrowa, singer_2006_narrowb, singer_2006_narrowc, schuss_2007_narrow} by focusing on the limiting case of small escape regions on the domain boundary. In two dimensions, the leading order of the MFPT is known to be proportional to the logarithm of harmonic measure of the escape region, which is proportional to its perimeter in the case of a disk-like domain~\cite{holcman_2004_escape, grebenkov_2016_universal}. Also the sub-leading behavior of the MFPT has been studied which plays an important role for the logarithmic dependence and depends generally on the pseudo Green's function of the domain~\cite{condamin_2007_firstpassage, benichou_2008_narrowescape, cheviakov_2010_asymptotic, pillay_2010_asymptotic, chevalier_2011_firstpassage, cheviakov_2012_mathematical}. An exact solution of the escape problem, via a general integral formula, was recently shown for any simply connected planar domain with an arbitrary diffusivity implying the conformal mapping of the unit disk onto this domain~\cite{grebenkov_2016_universal}. Moreover, some asymptotically exact formulas was found for non-spherical three-dimensional domains~\cite{gomez_2015_asymptotic}.

Recently the narrow escape problem was studied in spatially inhomogeneous environments~\cite{schwarz_2016_optimality}. The type of spatial inhomogeneity was inspired by the spatial organization of the cytoskeleton of cells with a centrosome~\cite{schwarz_2016_numerical, hafner_2016_spatial, hafner_2018_spatial} along which ballistic transport is possible in addition to simple diffusion: a circular or spherical domain is divided into two concentric shells, the inner shell allowing only radial ballistic transport (and diffusion) and the outer shell of width $\Delta$ allows multi-directional ballistic transport (and diffusion). In these studies it was shown that the MFPT for the narrow escape problem is optimizable with respect to the width of the outer shell. The physical reason for the optimization of the MFPT in the considered setup was argued to be related to the accelerated effective diffusion constant in the outer shell~\cite{schwarz_2016_optimality,schwarz_2016_numerical, hafner_2016_spatial, hafner_2018_spatial}, in reminiscence of the optimization of the reaction time in the presence of surface-mediated diffusion~\cite{benichou_2010_optimal, benichou_2011_mean, calandre_2012_interfacial, rupprecht_2012_kinetics, rupprecht_2012_exact} with different models of absorption and desorption.

In view of this reminiscence the natural question arises whether the MFPT also shows a non-monotonous - and hence optimizable - dependence on the thickness of the outer shell if ballistic transport is neglected and only the diffusion constant in the outer shell is larger than the diffusion constant in the inner shell. Therefore in this paper we consider this two-shell geometry with only a diffusive motion with two different diffusivities in the center of the cell and in the cortex of width $\Delta$. Diffusive motion in such discontinuous media, presenting a discontinuity of the coefficient of diffusion throughout an interface, is little studied. The one-dimensional profile of concentration of diffusive particles in such domain, presenting a jump close to the interface, was investigated~\cite{labolle_2000_diffusion, alvarez-ramirez_2014_asymmetric, alvarez-ramirez_2014_asymmetrical}. Moreover, a simulation scheme for one-dimensional problems is presented in the It\=o formalism taking into account the drift term proportional to the gradient of the diffusivity~\cite{lejay_2006_scheme, lejay_2012_simulating, lejay_2013_new} and compatible with the Fick's semi-empirical law. In fact, different writings of Langevin equation are possible in heterogeneous media, where the diffusivity is spatially varying~\cite{vaccario_2015_firstpassage}. 

\begin{figure}[t]
\begin{center}
  \includegraphics[width=16cm]{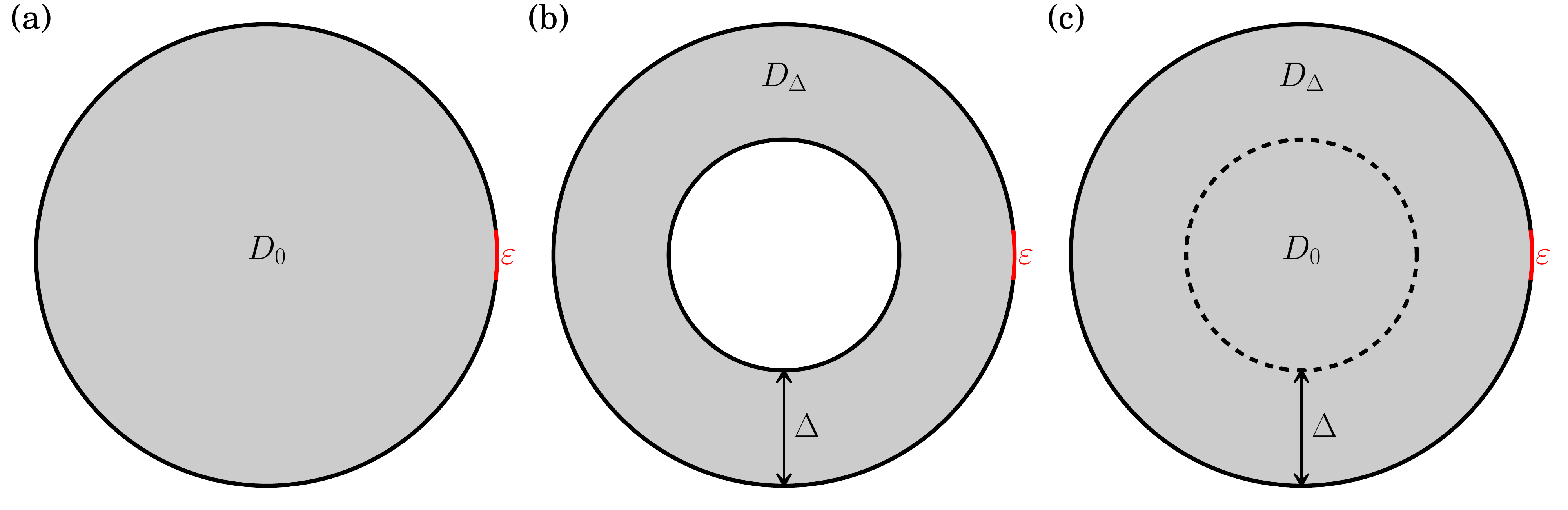}
  \caption{Different geometries studied here. The grey region represents the available space for the Brownian particle, denoted as $\Omega$. The escape region $\partial \Omega_\varepsilon$ of angle $\varepsilon$ is located on the outer circle denoted as $\partial \Omega$. For the two last geometries, the inner circle is denoted as $\partial \Omega_\Delta$ and the width of the outer shell is denoted as $\Delta$. {\bf (a)}~Disk geometry mentionned in section \ref{sec1}. {\bf (b)}~Annulus geometry studied in section \ref{sec2}. {\bf (c)}~Two-shell geometry studied in section \ref{sec3}. \label{fig_domains}}
\end{center}
\end{figure}

In this article, we will be interested in the first passage problem for the disk-like geometries of radius ${\cal R}$ presenting one target of angle $\varepsilon$ located on the outer boundary, which represents an opening of length $\varepsilon {\cal R}$. The domain is separated into two shells: the inner shell of radius ${\cal R}-\Delta$ playing the role of the center of the cell, where the diffusive particle has a diffusivity $D_0$ and the outer shell of width $\Delta$ playing the role of the cortex, where the diffusive particle has a diffusivity $D_\Delta$. Since our main motivation was to investigate an idealized case in which the outer shell represents the actin cortex of a cell and therefore has a higher effective diffusion constant we mainly, but not exclusively, look at $D_\Delta>D_0$. We will look at two different expansions within this circular discontinuous geometry: the narrow escape problem where the escape region is small ($\varepsilon \ll 1$) and the thin outer region problem ($\Delta \ll {\cal R}$). We present in section \ref{sec1} a well-known solution of the narrow escape problem of the disk (same diffusivity in the two shells), with a geometry represented in the Fig.~\ref{fig_domains}a. In section \ref{sec2}, we study the annulus geometry, shown in Fig.~\ref{fig_domains}b, where the diffusive particle is excluded from the inner shell. Finally, we will study in section \ref{sec3} the discontinuous problem whose the geometry is shown in Fig.~\ref{fig_domains}c. The section \ref{sec4} concludes with a discussion and an outlook.

\section{Narrow escape problem for the disk-like geometries}
\label{sec1}

We first look at the escape problem for the two-dimensional disk-like geometries. This problem is equivalent to the two-shell geometry with the same diffusivity in both shells such that $D_\Delta = D_0$ which is independent of the value of $\Delta$. The available space for the Brownian particles is a disk of radius ${\cal R}$, denoted as $\Omega$, and its area is equal to $|\Omega| = \pi {\cal R}^2$. The external boundary is located at the radius $|{\bf x}|={\cal R}$ and is denoted as $\partial \Omega$ whereas the absorbing part of this boundary located at $|\theta| \le \varepsilon/2$ is denoted as $\partial \Omega_\varepsilon$. The geometry of this domain is sketched in Fig.~\ref{fig_domains}a. The MFPT, denoted as $t({\bf x})$ for a Brownian particle starting at the point ${\bf x}$, can be related to the probability density function $p({\bf y},\tau|{\bf x})$ to be at the point ${\bf y}$ after a time $\tau$ without having reached the escape region via
\begin{equation}
\label{linkMFPTprob}
t({\bf x}) = \int_\Omega d{\bf y} \int_0^\infty d\tau \ p({\bf y},\tau|{\bf x}).
\end{equation}
The equations for the function $t({\bf x})$ are then deduced from the backward equations for the probability density function~\cite{redner_2001_guide}
\begin{gather}
D_0 \nabla^2 t({\bf x}) = -1, \quad {\bf x} \in \Omega \label{FPT1} \\
t({\bf x}) = 0, \quad {\bf x} \in \partial \Omega_\varepsilon \label{FPT2} \\
{\bf n} \cdot \nabla t({\bf x}) = 0, \quad {\bf x} \in \partial \Omega \backslash \partial \Omega_\varepsilon \label{FPT3}.
\end{gather}

\begin{figure}[t]
\begin{center}
  \includegraphics[width=16cm]{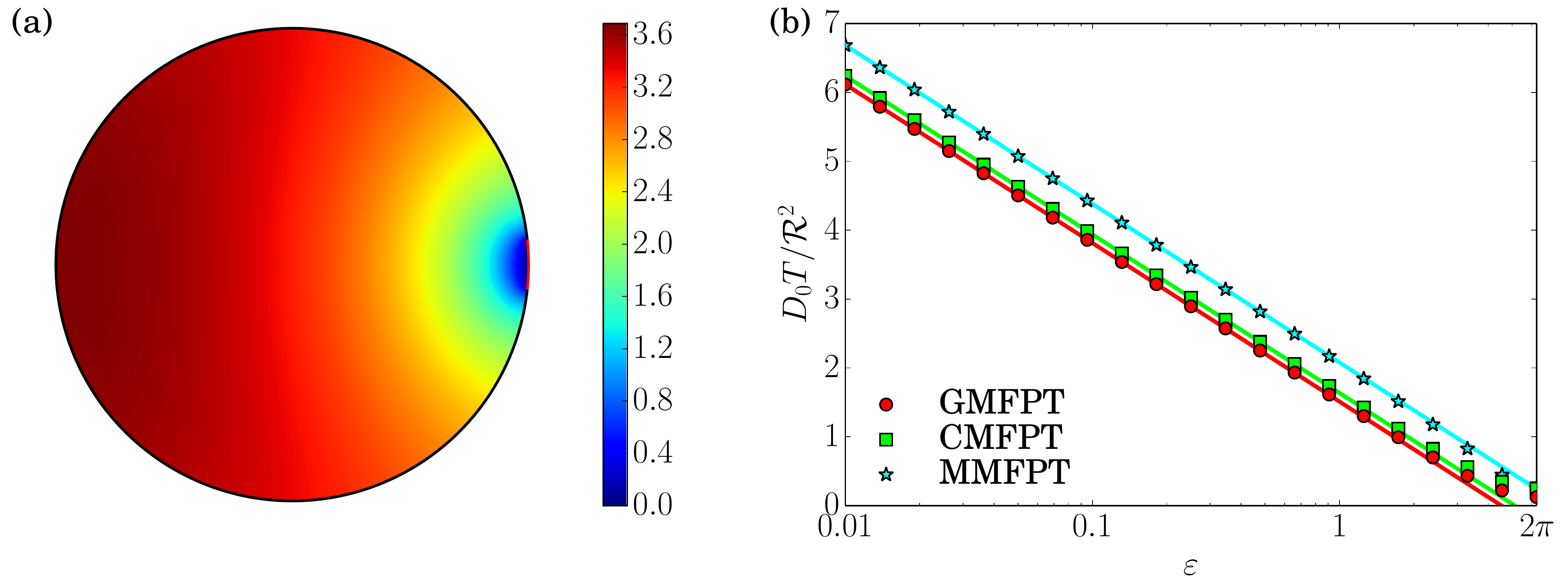}
  \caption{{\bf (a)}~Numerical solution of the dimensionless mean first passage time (MFPT) for a Brownian particle starting at ${\bf x}$: $D_0 t({\bf x})/{\cal R}^2$ for $\varepsilon=0.2$ obtained by solving the Eqs.~(\ref{FPT1}-\ref{FPT3}) with FreeFem++. {\bf (b)}~Dimensionless global MFPT (GMFPT): $D_0 \langle t \rangle /{\cal R}^2$, dimensionless MFPT for a particle starting at the center (CMFPT): $D_0 t({\bf 0})/{\cal R}^2$ and dimensionless maximal MFPT (MMFPT): $D_0 t_{\rm max}/{\cal R}^2$ plotted as a function of $\varepsilon$. The numerical solutions (symbols) are compared to the analytical solution (full lines) given by Eqs.~(\ref{circularGMFPT}),~(\ref{circularCMFPT}) and~(\ref{circularMMFPT}) respectively. These expressions obtained in the narrow escape limit are quite accurate for $\varepsilon \lesssim 2\pi/3$. The fitted value of the regular part of the pseudo Green's function close to the hole is $\pi R({\bf x_0}|{\bf x_0}) \simeq 0.124728$ which represents a relative error of $0.217\%$. \label{fig_circular2d}}
\end{center}
\end{figure}

The physical parameters of this problem are: the diffusivity of the Brownian particle $D_0$, the radius of the disk ${\cal R}$ and the angle $\varepsilon$ of the escape region. Only one pertinent parameter remains to study the narrow escape (NE) problem: $\varepsilon$, using dimensionless variables after rescaling the length by the radius ${\cal R}$ and the time by ${\cal R}^2/D_0$. The numerical solution of Eqs.~(\ref{FPT1}-\ref{FPT3}) for the MFPT obtained with the PDE solver FreeFem++~\cite{hecht_2013_new} is shown in Fig.~\ref{fig_circular2d}a for the escape angle $\varepsilon = 0.2$.

In this section, we will reproduce the main results of the literature~\cite{kolokolnikov_2005_optimizing, singer_2006_narrowb, schuss_2007_narrow, pillay_2010_asymptotic, chevalier_2011_firstpassage, cheviakov_2012_mathematical, caginalp_2012_analytical, rupprecht_2015_exit} for the MFPT of the disk geometry, which will constitute the starting point of our analysis. The leading order of the MFPT in the NE limit ($\varepsilon \ll 1$) is given by the constant solution equal to the spatial average value of the MFPT~\cite{holcman_2004_escape}
\begin{equation}
\label{leading}
t({\bf x}) \simeq \langle t \rangle \simeq -\frac{|\Omega|}{\pi D_0} \ln \varepsilon.
\end{equation}
Applying the Benichou's method~\cite{condamin_2007_firstpassage, benichou_2008_narrowescape}, one obtains straightforwardly the space-dependent solution in leading order
\begin{equation}
\label{Benichou}
t({\bf x}) \simeq \frac{|\Omega|}{\pi D_0} \ln \frac{|{\bf x} - {\bf x_0}|}{\varepsilon {\cal R}},
\end{equation}
where ${\bf x_0} = ({\cal R},0)$ is the center of the escape region. However, the space average of the MFPT remains equal to Eq.~(\ref{leading}) since the average value of $\ln(|{\bf x} - {\bf x_0}|/{\cal R})$ is zero over the disk. To obtain the order ${\cal O}(\varepsilon^0)$ result for the MFPT, which is important even if $\varepsilon$ is as small as $10^{-2}$ due to the logarithmic behavior, we introduce the pseudo Green's function $G({\bf x}|{\bf x_0})$, following Refs.~\cite{pillay_2010_asymptotic, chevalier_2011_firstpassage}, defined by the equations
\begin{gather}
\nabla^2 G({\bf x}|{\bf x_0}) = \frac{1}{|\Omega|}, \quad {\bf x} \in \Omega \label{Green1}\\
{\bf n} \cdot \nabla G({\bf x}|{\bf x_0}) = 0 , \quad {\bf x} \in \partial \Omega \label{Green2}\\
G({\bf x} \rightarrow {\bf x_0} |{\bf x_0}) = -\frac{1}{\pi} \ln \frac{|{\bf x} - {\bf x_0}|}{{\cal R}} + R({\bf x_0}| {\bf x_0}), \label{Green3}\\
\int_{\Omega} d{\bf x}\ G({\bf x}|{\bf x_0}) = 0 \label{Green4},
\end{gather}
where $R({\bf x_0}| {\bf x_0})$ is the (unknown) regular part of the Green's function at the center of the escape region ${\bf x_0}$. The spatial average of the MFPT, hereinafter called Global Mean First Passage Time (GMFPT), is given by
\begin{equation}
\label{expGMFPT}
\langle t \rangle =  \frac{|\Omega|}{\pi D_0} \left[ - \ln \frac{\varepsilon}{4} + \pi R({\bf x_0}|{\bf x_0}) \right],
\end{equation}
and the expression of the MFPT $t({\bf x})$ is
\begin{equation}
\label{expMFPTlitt}
t({\bf x}) = \langle t \rangle - \frac{|\Omega|}{D_0} G({\bf x}|{\bf x_0}).
\end{equation}
A general proof of these equations is shown in the appendix \ref{appendixNE}. The equations (\ref{Green1}-\ref{Green4}) can be solved for the disk-like geometries we study here, for the regular part of the Green's function defined by
\begin{equation}
\label{defofR}
R({\bf x}|{\bf x_0}) = G({\bf x}|{\bf x_0}) + \frac{1}{\pi} \ln \frac{|{\bf x} - {\bf x_0}|}{{\cal R}}.
\end{equation}
The MFPT function $t({\bf x})$ satisfies thus the expression
\begin{equation}
\label{expMFPT}
t({\bf x}) = \frac{|\Omega|}{\pi D_0}  \left[ \ln \frac{4 |{\bf x} - {\bf x_0}|}{\varepsilon {\cal R}} + \pi R({\bf x_0}|{\bf x_0}) - \pi R({\bf x}|{\bf x_0}) \right].
\end{equation}

For the disk geometry represented in Fig.~\ref{fig_domains}, the regular part of the Green's function satisfies the Eq.~(\ref{Rdisk}). Its value close to the escape region is equal to $R({\bf x_0}|{\bf x_0}) = 1/8\pi$. The GMFPT expression is
\begin{equation}
\label{circularGMFPT}
\langle t \rangle \underset{\varepsilon \ll 1}{\simeq}  \frac{{\cal R}^2}{D_0} \left[ - \ln \frac{\varepsilon}{4} + \frac{1}{8} \right],
\end{equation}
and the expression of the MFPT for a Brownian particle starting at ${\bf x}$ is
\begin{equation}
\label{circularMFPT}
t({\bf x}) \underset{\varepsilon \ll 1}{\simeq} \frac{{\cal R}^2}{D_0} \left[ \ln \frac{4 |{\bf x} - {\bf x_0}|}{\varepsilon {\cal R}} + \frac{{\cal R}^2-|{\bf x}|^2}{4{\cal R}^2} \right].
\end{equation}
At the leading order, this solution is comparable to Benichou's solution (\ref{Benichou}) close to the escape region. The expression of the MFPT for a Brownian particle starting at the center of the disk ${\bf x}= {\bf 0}$, denoted hereinafter as CMFPT, is then
\begin{equation}
\label{circularCMFPT}
t({\bf 0}) \underset{\varepsilon \ll 1}{\simeq} \frac{{\cal R}^2}{D_0} \left[ - \ln \frac{\varepsilon}{4} + \frac{1}{4} \right],
\end{equation}
and the expression of the maximal value of the MFPT, denoted hereinafter as MMFPT, is
\begin{equation}
\label{circularMMFPT}
t_{\rm max} \underset{\varepsilon \ll 1}{\simeq} t(-{\bf x_0}) \underset{\varepsilon \ll 1}{\simeq} \frac{{\cal R}^2}{D_0} \left[ - \ln \frac{\varepsilon}{4} + \ln 2 \right],
\end{equation}
corresponding to the value of the MFPT for a particle starting at the position $-{\bf x_0} = (-{\cal R},0)$, at the maximal distance to the escape region. The GMFPT, CMFPT and MMFPT values are shown in Fig.~\ref{fig_circular2d}b as a function of $\varepsilon$. The narrow escape expressions given by Eqs.~(\ref{circularGMFPT}),~(\ref{circularCMFPT}) and~(\ref{circularMMFPT}) are compatible with the numerical solutions of the PDEs (\ref{FPT1}-\ref{FPT3}) for $\varepsilon \lesssim 2\pi/3$.

These numerical solutions are obtained with the finite element method using the software package FreeFem++~\cite{hecht_2013_new}, with a relative interpolation error of order $10^{-4}$ for a mesh grid built with around $10^4$ vertices. For our study of the GMFPT, we can compare the numerically interpolated value of $\pi R({\bf x_0}|{\bf x_0})$ which is approximatively equal to $0.124728$ with the analytical solution $1/8$ which represents a relative error of $0.217\%$. In the following sections, we considerer the data generated by FreeFem++ as the {\it exact} numerical solution.

Furthermore, Singer \etal \cite{singer_2006_narrowb} solved directly the partial differential equations for the MFPT (\ref{FPT1}-\ref{FPT3}) and gave an exact expression of the MFPT for all values of $\varepsilon$, as an infinite series with implicit coefficients written as an integral. The explicit form of these coefficients was given by Rupprecht \etal \cite{rupprecht_2015_exit} in terms of Legendre polynomials. Finally, Caginalp and Chen \cite{caginalp_2012_analytical} derived a fully explicit closed formula for the mean first-passage time which writes
\begin{equation}
\label{exactDisk}
\frac{D_0 t(z)}{{\cal R}^2} = \frac{1-|z|^2}{4} + \ln \left|\frac{1-z+\sqrt{[1-z\exp(-i\varepsilon/2)][1-z\exp(i\varepsilon/2)]}}{2\sin(\varepsilon/4)}\right|
\end{equation}
with respect to the complex variable $z=r\exp(i\theta)$ such that $r=|{\bf x}|/{\cal R}$ and $\theta$ represents the angle between ${\bf x}$ and the center of the escape region ${\bf x_0}$. This exact expression is asymptotically compatible with Eq.~(\ref{circularMFPT}) in the $\varepsilon \ll 1$ limit. We also mention here that the result of Rupprecht \etal \cite{rupprecht_2015_exit} provides the full distribution of the first passage times.

To solve the annulus and two-shell geometry problems, we will use the same formalism to solve the equations for the regular part of the pseudo Green's function $R({\bf x}|{\bf x_0})$. Then we apply Eqs.~(\ref{expGMFPT}) and (\ref{expMFPT}) to obtain the expressions of the GMFPT and the MFPT for a particle starting at the position ${\bf x}$, which are correct in the first sub-leading order of the small escape angle $\varepsilon$ expansion. In addition, we can mention that the recent exact solution of the escape problem for any simply connected planar domain with an arbitrary diffusivity~\cite{grebenkov_2016_universal}. This exact solution is compatible with the Eq.~(\ref{exactDisk}) for the disk geometry. 

%%% SECTION 2

\section{The narrow escape problem for the annulus geometry}
\label{sec2}

\begin{figure}[t]
\begin{center}
  \includegraphics[width=16cm]{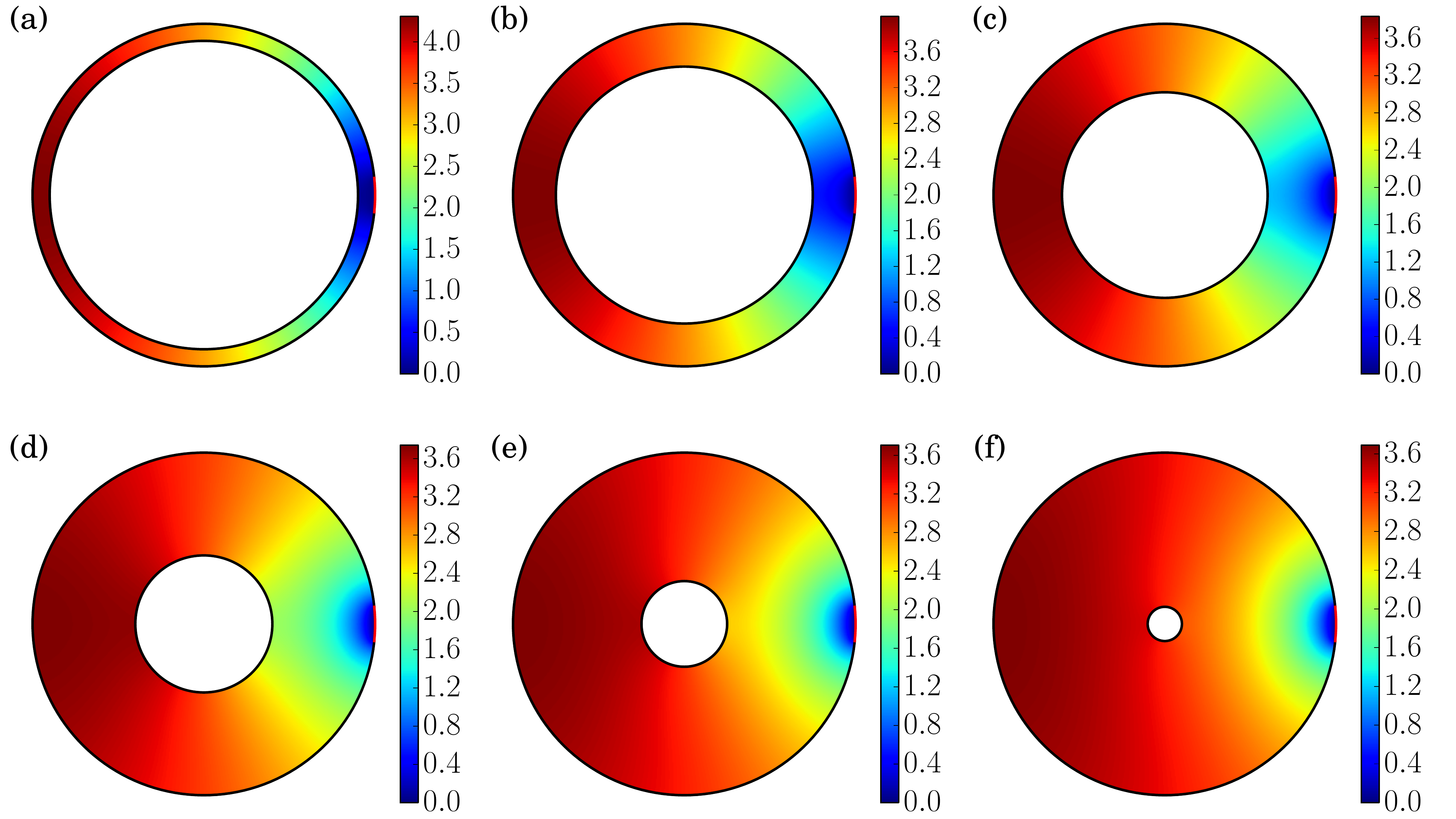}
  \caption{Numerical solution of the dimensionless mean first passage time $D_\Delta t({\bf x})/{\cal R}^2$ for a particle starting at ${\bf x}$ with an escape angle $\varepsilon=0.2$ obtained by solving the Eqs.~(\ref{FPTannulus1}-\ref{FPTannulus4}) with FreeFem++ for several values of $\Delta$: {\bf (a)}~$\Delta/{\cal R} = 0.1$, {\bf (b)}~$\Delta/{\cal R} = 0.25$, {\bf (c)}~$\Delta/{\cal R} = 0.4$, {\bf (d)}~$\Delta/{\cal R} = 0.6$, {\bf (e)}~$\Delta/{\cal R} = 0.75$ and {\bf (f)}~$\Delta/{\cal R} = 0.9$. The maximal value of $t({\bf x})$ is always located at $(-{\cal R},0)$. \label{fig_annulus_FPT}}
\end{center}
\end{figure}

In this section, we look at the NE problem for a Brownian particle inside an annulus of width $\Delta$, ${\cal R}-\Delta \le |{\bf x}| \le {\cal R}$, denoted as $\Omega$, with an area $|\Omega| = \pi \Delta (2{\cal R}-\Delta)$. The outer boundary $|{\bf x}|={\cal R}$ is denoted as $\partial \Omega$ while the inner boundary $|{\bf x}| = {\cal R}-\Delta$ is denoted as $\partial \Omega_\Delta$. The escape region is part of the outer boundary defined by $|\theta| \le \varepsilon /2$ and is denoted as $\partial \Omega_\varepsilon$. The opposite case with an escape region on the inner boundary was studied by Holcman \etal \cite{schuss_2007_narrow}. This problem is equivalent to the two-shell geometry in the limit case $D_0=0$ by which the diffusive particle is excluded from the inner shell ($|{\bf x}|\le {\cal R}-\Delta$). The equations for the MFPT for a Brownian particle starting at the position ${\bf x}$, denoted as $t({\bf x})$, are deduced from the Eq.~(\ref{linkMFPTprob}) and the backward equations for the probability density function~\cite{redner_2001_guide}
\begin{gather}
D_\Delta \nabla^2 t({\bf x}) = -1, \quad {\bf x} \in \Omega \label{FPTannulus1} \\
t({\bf x}) = 0, \quad {\bf x} \in \partial \Omega_\varepsilon \label{FPTannulus2} \\
{\bf n} \cdot \nabla t({\bf x}) = 0 , \quad {\bf x} \in \partial \Omega \backslash \partial \Omega_\varepsilon \label{FPTannulus3}\\
{\bf n} \cdot \nabla t({\bf x}) = 0 , \quad {\bf x} \in \partial \Omega_\Delta. \label{FPTannulus4}
\end{gather}
This problem can be solved with only two dimensionless parameters: the dimensionless annulus width $\delta = \Delta/{\cal R}$ with respect to the radius ${\cal R}$ and the angle $\varepsilon$ of the escape region, after rescaling the length by the radius ${\cal R}$ and the time by ${\cal R}^2/D_\Delta$. Numerical solutions of these PDEs are shown in the Fig.~\ref{fig_annulus_FPT} for several values of the annulus width $\Delta \in [0,{\cal R}]$ and the escape angle $\varepsilon=0.2$. When $\Delta \rightarrow {\cal R}$, the solutions of the section \ref{sec1} are valid, replacing $D_0$ by $D_\Delta$ in analytical expressions, comparing the Fig.~\ref{fig_annulus_FPT}f with the Fig.~\ref{fig_circular2d}a.

Fig.~\ref{fig_annulus_FPT} demonstrates several interesting features of the MFPT: by comparing (a) with (f) one sees that particles starting close to the escape region find the target faster in the thin annulus (a), {\it i. e.} small $\Delta$, than in the thick annulus (f), {\it i. e.} large $\Delta$. The inaccessible central region represents an obstacle impeding the particles starting close to the escape region from diffusing away when it is sufficiently large. On the other hand particles starting far away from the escape region, for instance on the opposite side, reach the target faster in the thick annulus (f), {\it i. e.} large $\Delta$, than in the thin annulus (a), {\it i. e.} small $\Delta$. Those particle are impeded by the central obstacle to move quickly towards the target, they remain {\it jammed} for a longer time when the obstacle is large. Thus the MFPT for particles starting close to the escape region increases with increasing $\Delta$, whereas the MFPT of particles starting far away from the escape region decreases with decreasing $\Delta$. After averaging over the initial positions these two opposing tendencies could in principle lead to a minimum of the GMFPT as function of $\Delta$. This is actually what we will find in the next subsection.

\subsection{The narrow escape limit ($\varepsilon \ll 1$)}

\begin{figure}[t]
\begin{center}
  \includegraphics[width=8cm]{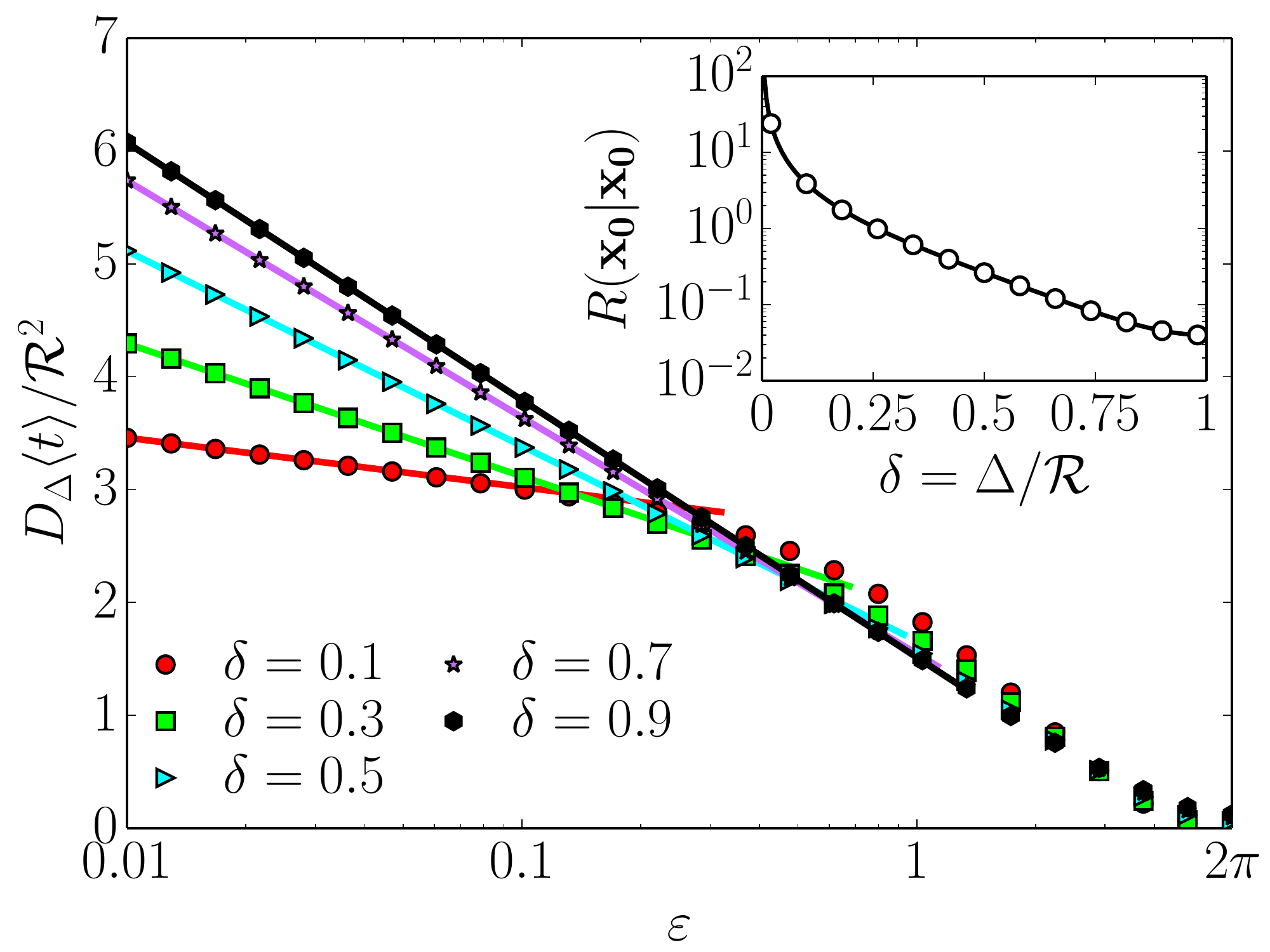}
  \caption{Numerical solution (symbols) of the dimensionless global mean first passage time $D_\Delta \langle t \rangle/{\cal R}^2$ compared with the analytical expression (straight lines) given by Eq.~(\ref{GMFPTAnnulus}) for several values of $\Delta$. In inset, the fitted value of $R({\bf x_0}|{\bf x_0})$, obtain from Eq.~(\ref{expGMFPT}), is plotted as a function of $\delta=\Delta/{\cal R}$ and compared to its analytical expression~(\ref{RGreenAnnulus}). This function goes to $1/8\pi$ for $\Delta \rightarrow {\cal R}$ and diverges as $\pi {\cal R}/6\Delta$ for $\Delta \rightarrow 0$. \label{fig_annulus_NE}}
\end{center}
\end{figure}

The derivation of the sub-leading order of the MFPT is presented in appendix \ref{appendixNE}, where it is shown that the MFPT satisfies Eq.~(\ref{expMFPT}) with the volume replaced by $|\Omega| = \pi \Delta (2{\cal R}-\Delta)$ and $D_0$ by $D_\Delta$. The regular part of the Green's function $R({\bf x}| {\bf x_0})$ is calculated in the appendix \ref{appendixAnnulus} and its expression is given by Eq.~(\ref{Rannulus}) in terms of (dimensionless) polar coordinates. Its value at the center of the escape region is then equal to
\begin{equation}
\label{RGreenAnnulus}
R({\bf x_0}|{\bf x_0}) = \frac{3}{8\pi} - \frac{1}{4\pi \delta(2-\delta)} -\frac{(1-\delta)^4 \ln(1-\delta)}{2\pi\delta^2(2-\delta)^2} + \frac{2}{\pi} \sum_{n=1}^\infty \frac{(1-\delta)^{2n}}{n\left[1-(1-\delta)^{2n}\right]},
\end{equation}
which implies the GMFPT expression from Eq.~(\ref{expGMFPT})
\begin{equation}
\label{GMFPTAnnulus}
\frac{D_\Delta \langle t \rangle}{{\cal R}^2} \underset{\varepsilon \ll 1}{\simeq}  \delta(2-\delta) \left[ - \ln \frac{\varepsilon}{4} + \frac{3}{8}  + 2 \sum_{n=1}^\infty \frac{(1-\delta)^{2n}}{n\left[1-(1-\delta)^{2n}\right]} \right] - \frac{1}{4} -\frac{(1-\delta)^4 \ln(1-\delta)}{2\delta(2-\delta)}.
\end{equation}
In Fig.~\ref{fig_annulus_NE}, the GMFPT is plotted as a function of the escape angle $\varepsilon$ for several annulus widths $\Delta \in [0,{\cal R}]$. In the limit $\varepsilon \ll 1$, we observe that the analytical prediction given by Eq.~(\ref{GMFPTAnnulus}) is verified in comparison with the numerical solutions obtained with FreeFem++. Moreover, we remark in the inset of the Fig.~\ref{fig_annulus_NE} that the value of the regular part of the Green's function at ${\bf x_0}$ plotted with respect to the dimensionless annulus width $\delta$, extract from numerical solutions, is compatible with the Eq.~(\ref{RGreenAnnulus}).

The value of $R({\bf x_0}|{\bf x_0})$ depends on the function $S(X)$ defined by the series expression (for $|X|< 1$)
\begin{equation}
S(X) = \sum_{n=1}^\infty \frac{X^{2n}}{n(1-X^{2n})}.
\end{equation}
When $X \rightarrow 0$, this function behaves like $S(X) \simeq - \ln (1-X^2)$. Then, in the limit $\delta \rightarrow 1$, the regular part of the pseudo Green's function behaves like $R({\bf x_0}|{\bf x_0}) \sim 1/8\pi$, corresponding to the result of the section \ref{sec1} which implies that the GMFPT satisfies the Eq.~(\ref{circularGMFPT}). When $X \rightarrow 1$, the leading order of the series is $S(X) \simeq \pi^2 (1-X)^{-1}/12$. Thus, in the limit $\Delta \rightarrow 0$, the regular part of the pseudo Green's function behaves like $R({\bf x_0}|{\bf x_0}) \sim \pi {\cal R} / 6\Delta$ which implies the leading order of the GMFPT in the $\varepsilon \rightarrow 0$ and $\delta \rightarrow 0$ limits
\begin{equation}
\label{GMFPTAnnulus00}
\frac{D_\Delta \langle t \rangle}{{\cal R}^2} \simeq -2\delta \ln \frac{\varepsilon}{4} + \frac{\pi^2}{3} + {\cal O}(\delta,\varepsilon).
\end{equation}

When $\varepsilon=2\pi$, the mixed Neumann-Dirichlet boundary condition becomes a fully absorbing (Dirichlet) boundary condition. The MFPT for a particle starting at ${\bf x}$ is then radially symmetric and the GMFPT expression is
\begin{equation}
\label{GMFPTAnnulusFull}
\frac{D_\Delta \langle t \rangle}{{\cal R}^2} \underset{\varepsilon = 2 \pi}{=}  \frac{3 \delta(2-\delta) }{8} - \frac{1}{4} -\frac{(1-\delta)^4 \ln(1-\delta)}{2\delta(2-\delta)}.
\end{equation}
The expression (\ref{GMFPTAnnulus}) of the GMFPT in the narrow escape limit can be decomposed as the sum of the GMFPT to reach the boundary $\partial \Omega$ given by Eq.~(\ref{GMFPTAnnulusFull}) and the GMFPT starting at the boundary $\partial \Omega$ to reach the escape region which writes in the NE regime as 
\begin{equation}
\frac{D_\Delta}{{\cal R}^2} t(\partial \Omega \rightarrow \partial \Omega_\varepsilon) \simeq \delta(2-\delta) \left[ - \ln \frac{\varepsilon}{4} + 2 \sum_{n=1}^\infty \frac{(1-\delta)^{2n}}{n\left[1-(1-\delta)^{2n}\right]} \right].
\end{equation}

Moreover, the maximal value of the MFPT is located in the NE regime at $-{\bf x_0}=(-{\cal R},0)$. Its expression is then obtained from Eqs.~(\ref{expMFPT}) and~(\ref{Rannulus}) with ${\bf x}=-{\bf x_0}$
\begin{equation}
\label{MMFPTAnnulus}
\frac{D_\Delta t_{\rm max}}{{\cal R}^2}  \simeq \frac{D_\Delta t(-{\bf x_0})}{{\cal R}^2} \underset{\varepsilon \ll 1}{\simeq}  \delta(2-\delta) \left[ - \ln \frac{\varepsilon}{8} + 4 \sum_{k=0}^\infty \frac{(1-\delta)^{2(2k+1)}}{(2k+1)\left[1-(1-\delta)^{2(2k+1)}\right]} \right].
\end{equation}
In the $\varepsilon \rightarrow 0$ and $\delta \rightarrow 0$ limits, this expression of the MMFPT goes to the value
\begin{equation}
\label{MMFPTAnnulus00}
\frac{D_\Delta t_{\rm max}}{{\cal R}^2} \simeq -2 \delta \ln \frac{\varepsilon}{8} + \frac{\pi^2}{2} + {\cal O}(\delta,\varepsilon).
\end{equation}

In the opposite regime ($\varepsilon \simeq 2\pi $), the maximal value of the MFPT is located on the circle of radius $|{\bf x}|={\cal R}-\Delta$ and its expression is then given by
\begin{equation}
\label{MMFPTAnnulusFull}
\frac{D_\Delta t_{\rm max}}{{\cal R}^2} \underset{\varepsilon = 2 \pi}{=}  \frac{\delta(2-\delta)}{4} + \frac{(1-\delta) \ln(1-\delta)}{2}
\end{equation}
for a fully absorbing boundary.

\begin{figure}[t]
\begin{center}
  \includegraphics[width=16cm]{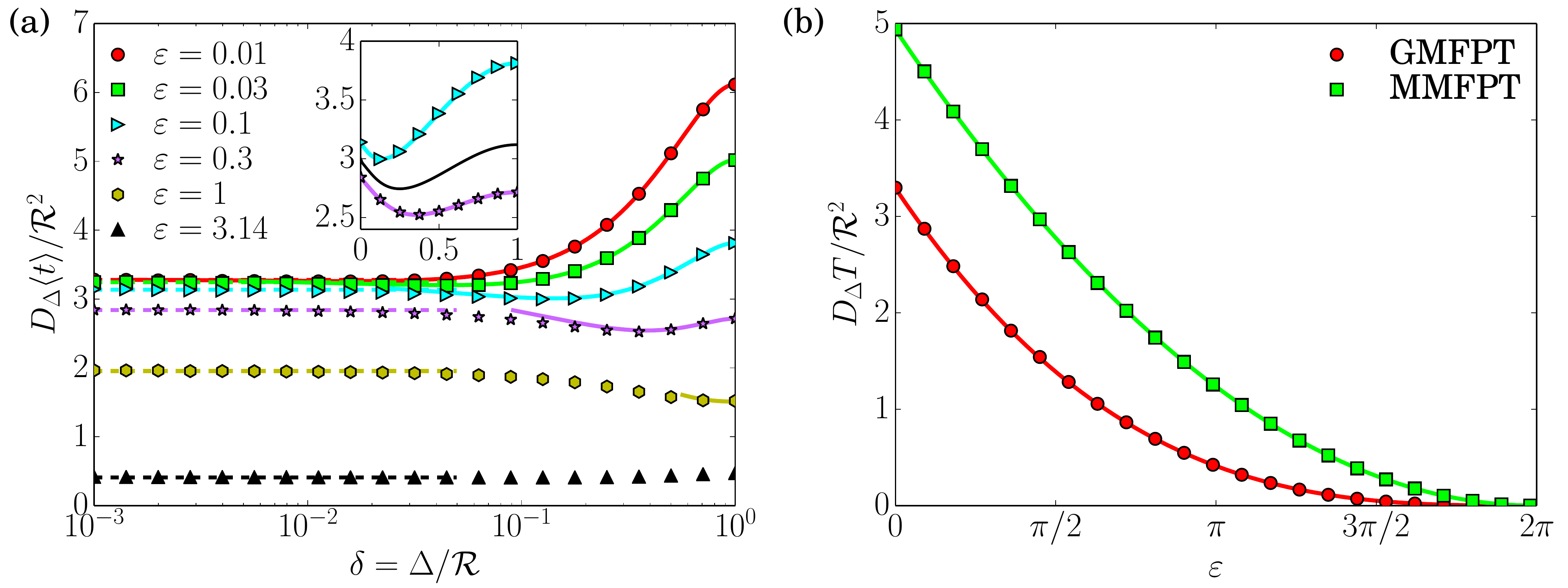}
  \caption{{\bf (a)}~Numerical solution (symbols) of the dimensionless global mean first passage time (GMFPT) $D_\Delta \langle t \rangle/{\cal R}^2$ as a function of $\delta=\Delta/{\cal R}$ for several values of $\varepsilon$. It is compared respectively with straight and dashed lines to the narrow escape expression (\ref{GMFPTAnnulus}) for $\varepsilon \ll1$ and $- \delta \ln\varepsilon \gg 1$ and the pseudo one-dimensional solution (\ref{expAnnulusGMFPT}) for $\delta \ll 1$ and $- \delta \ln\varepsilon \ll 1$. The GMFPT presents a minimum for fixed value of $\varepsilon<1$ at $0 < \Delta \le {\cal R}$, well-described by the narrow escape expression. In the inset, the presence of this minimum is represented for $\varepsilon \in\{0.1,0.2,0.3\}$ for a linear scale. {\bf (b)}~Numerical solution of the dimensionless GMFPT (circles) and MMFPT (squares) as a function of $\varepsilon \ge 10^{-2}$ for $\delta = 10^{-3}$ compared to the analytical solution given respectively by Eq.~(\ref{expAnnulusGMFPT}) and Eq.~(\ref{expAnnulusMMFPT}) obtained in the limit $\Delta \ll {\cal R}$ (straight lines). \label{fig_annulus_1d}}
\end{center}
\end{figure}

\begin{figure}[t]
\begin{center}
  \includegraphics[width=16cm]{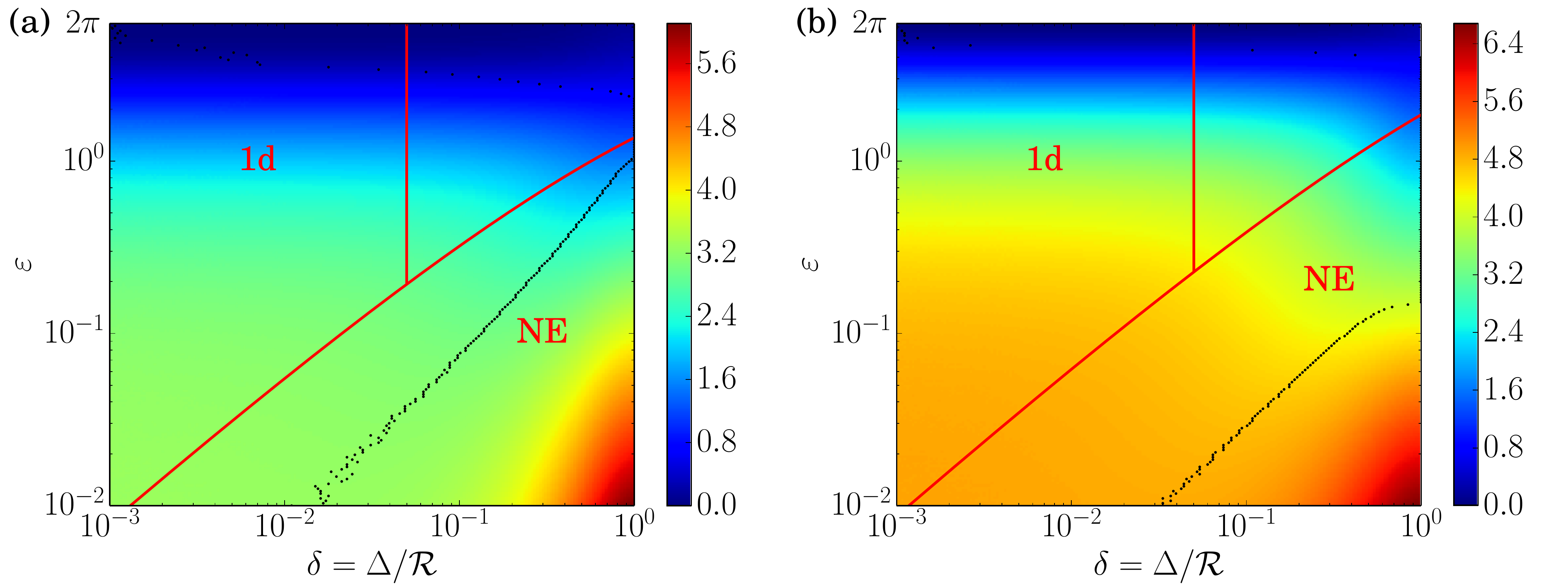}
  \caption{Numerical solution of the dimensionless global mean first passage time (GMFPT) $D_\Delta \langle t \rangle/{\cal R}^2$~{\bf (a)} and maximal mean first passage time (MMFPT) $D_\Delta t_{\rm max}/{\cal R}^2$~{\bf (b)} as functions of the annulus width $\Delta/{\cal R}$ and the escape angle $\varepsilon$. The black dots indicate the location of the minimum of the GMFPT and MMFPT as a function of $\Delta$ for fixed $\varepsilon$. The validity domain of the narrow escape (NE) expression ($\varepsilon \ll 1$ and $-\delta \ln \varepsilon > 1$) and the pseudo one-dimensional (1d) expression ($\delta \ll 1$ and $-\delta \ln \varepsilon \ll 1$) are represented in both figures.  \label{fig_annulus_minimum}}
\end{center}
\end{figure}

The GMFPT calculated by applying the PDE solver FreeFem++ to Eqs.~(\ref{FPTannulus1}-\ref{FPTannulus4}) is shown in Fig.~\ref{fig_annulus_1d}a as a function of the width of the annulus $\Delta$ for a fixed value of $\varepsilon$. One observes that the GMFPT approaches a constant in the thin annulus limit. Furthermore, the GMFPT is minimized for a specific value of the width $\Delta$ of the annulus. For an escape angle $\varepsilon \lesssim 1$, the minimum is pronounced and clearly visible in Fig.~\ref{fig_annulus_1d}a, whereas for smaller values of $\varepsilon$ the minimum is very shallow and nearly invisible but indicated by the black dots in Fig.~\ref{fig_annulus_minimum}.

In connection with Fig.~\ref{fig_annulus_FPT} we have already realized that the MFPT for particles starting close to the escape region increases with increasing $\Delta$, whereas the MFPT of particles starting far away from the escape region decreases with decreasing $\Delta$ such that these opposing tendencies could lead to a minimum of the GMFPT as a function of $\Delta$. This picture is confirmed by the behavior of the GMFPT presented in Fig.~\ref{fig_annulus_minimum}a, where the black dots indicate the value of~$\Delta$ where for fixed $\varepsilon$ the minimum occurs. The starting point with the maximum MFPT is always $-{\bf x_0} = (-{\cal R},0)$ and Fig.~\ref{fig_annulus_minimum}b shows clearly that for $\varepsilon<0.15$ the MFPT of a particle starting opposite to the escape region first decreases with increasing $\Delta$, reaches a minimum at a specific width and then increases with $\Delta$.

\begin{figure}[t]
\begin{center}
  \includegraphics[width=16cm]{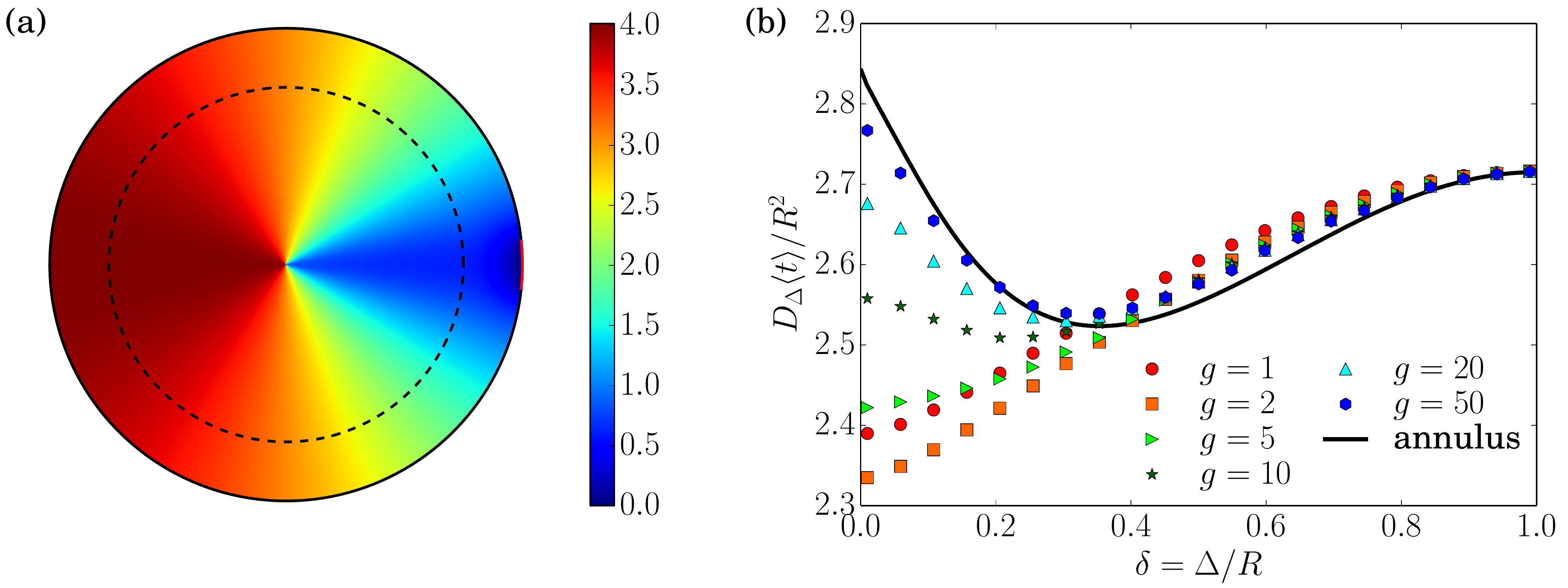}
  \caption{{\bf (a)}~Numerical solution of the dimensionless mean first passage time $D_\Delta t({\bf x})/{\cal R}^2$ for a particle starting at ${\bf x}$ for a width $\Delta=0.25{\cal R}$, an escape angle $\varepsilon=0.2$ obtained by solving the Eqs.~(\ref{FPTsoft1}-\ref{FPTsoft3}) for the potential (\ref{logarithmicPot}) with $g=100$. {\bf (b)}~Numerical solution (symbols) of the dimensionless global mean first passage time (GMFPT) $D_\Delta \langle t \rangle/{\cal R}^2$ as a function of $\delta=\Delta/{\cal R}$ for $\varepsilon=0.3$ and several values of $g$ compared to the numerical solution obtained for the annulus geometry (see inset of Fig.\ref{fig_annulus_1d}a). Note that the minimum appears for $g\ge 10$. \label{fig_annulus_soft}}
\end{center}
\end{figure}

It is an interesting question whether the existence of the minimum discussed here persists when the impenetrable central obstacle, $|{\bf x}|<{\cal R}-\Delta$, is replaced by a {\it soft} repulsive barrier, which may occur because of geometric constraints in a cell. This situation is modeled by a Bessel process~\cite{ryabov_2015_brownian} which is an overdamped Brownian motion in a logarithmic potential $U({\bf x})$ for the inverse temperature $\beta$
\begin{equation}
\label{logarithmicPot}
\beta U({\bf x}) = -g \ln \frac{|{\bf x}|}{{\cal R}} \Theta({\cal R}-\Delta-|{\bf x}|)
\end{equation}
repulsive when $g>0$ and presenting a divergence at the center of the disk. $\Theta$ denotes here the Heaviside function. The MFPT starting at ${\bf x}$ to reach the escape region of angle $\varepsilon$ satisfies then the equations
\begin{gather}
D_\Delta \nabla^2 t({\bf x})-D_\Delta\beta  \nabla U({\bf x}) \cdot \nabla t({\bf x})  = -1, \quad {\bf x} \in \Omega \label{FPTsoft1} \\
t({\bf x}) = 0, \quad {\bf x} \in \partial \Omega_\varepsilon \label{FPTsoft2} \\
{\bf n} \cdot \nabla t({\bf x}) = 0, \quad {\bf x} \in \partial \Omega \backslash \partial \Omega_\varepsilon \label{FPTsoft3}.
\end{gather}
The boundary condition on $|{\bf x}|={\cal R}-\Delta$ is given by the conservation of the {\it backward} flux: $-D_\Delta {\bf n} \cdot \nabla t({\bf x})$ which does not depend on the value of the potential. The gradient of potential is
\begin{equation}
\beta \nabla U({\bf x}) = -g \frac{{\bf x}}{|{\bf x}|^2} \Theta({\cal R}-\Delta-|{\bf x}|).
\end{equation}
When $g=0$, the MFPT of the disk geometry is recovered (data not shown). In Fig.~\ref{fig_annulus_soft}a, the numerical solution MFPT obtained with the PDE solver FreeFem++ is shown for the width $\Delta=0.25{\cal R}$, the escape angle $\varepsilon=0.2$ and the potential strength $g=100$. The solution in the inner shell $|{\bf x}| < {\cal R} - \Delta$ is only $\theta$-dependent since the Brownian particle moves rapidly to the outer shell following the radial potential, equivalent to the boundary condition ${\bf n } \cdot \nabla t =0$ on the inner boundary. This solution is compatible with the annulus geometry in the outer shell (see Fig.~\ref{fig_annulus_FPT}b). A difference appears only when the potential is too {\it soft}, for example $g<10$, since MFPT is considerably changed in the limit $\Delta \ll {\cal R}$.

In Fig.~\ref{fig_annulus_soft}b, the GMFPT is shown as a function of the width $\Delta$ for $\varepsilon=0.3$ and several values of $g$ and compared to the GMFPT obtained for the annulus geometry. For $g=1$, the GMFPT of the annulus geometry is compatible only for $\Delta > 0.9 {\cal R}$ and the minimum is reached for $\Delta=0$ instead of $\Delta \simeq 0.4 {\cal R}$. A minimum exists for a width $\Delta \ne 0$ for $g \gtrsim 10$ and for $g>50$ the $\Delta$-dependance of the GMFPT is similar to the one of the annulus geometry. The minimum of the GMFPT is then not present for all {\it soft} potentials and a sufficiently strong repulsion is needed for it to appear. The lower limit of $g$, denoted as $g_l$, for the existence of a minimum depends on the escape angle $\varepsilon$ (data not shown) and decreases with $\varepsilon$.. In the narrow escape regime ($\varepsilon \ll 1$), this limit value is then large $g_l \gg 1$, which implies that the minimum appears only for a strong potential, like in the annulus geometry.

\subsection{The pseudo one-dimensional limit ($\Delta \ll {\cal R}$)}

In the $\Delta \ll {\cal R}$ limit, we can consider a pseudo one-dimensional diffusion on the surface $|{\bf x}|={\cal R}$ and Eqs.~(\ref{FPTannulus1}-\ref{FPTannulus2}) become for the function $t(\theta)$ dependent only on the orthoradial coordinate 
\begin{gather}
\frac{D_\Delta}{{\cal R}^2} t''(\theta) = -1 \\
t(\varepsilon/2) = t(2\pi - \varepsilon/2) = 0,
\end{gather}
while the Eqs.~(\ref{FPTannulus3}-\ref{FPTannulus4}) are automatically satisfied. The unique solution of this Cauchy system is
\begin{equation}
\label{expAnnulusFPTSurf}
\frac{D_\Delta}{{\cal R}^2} t(\theta) = \frac{(\theta-\varepsilon/2)(2\pi - \varepsilon/2 - \theta)}{2}.
\end{equation}

The GMFPT expression is then given by the spatial average over $\theta \in [0,2\pi]$ of the last expression
\begin{equation}
\label{expAnnulusGMFPT}
\frac{D_\Delta \langle t \rangle}{{\cal R}^2}  \underset{\Delta \ll {\cal R}}{\simeq}  \frac{(2\pi-\varepsilon)^3}{24\pi}.
\end{equation}
This expression reproduces the asymptotic value $\pi^2/3$ of the GMFPT as predicted by Eq.~(\ref{GMFPTAnnulus00}) in the limit $\varepsilon \rightarrow 0$ and $\delta \rightarrow 0$ simultaneous under the condition $-\delta \ln \varepsilon \ll 1$. The limit $\varepsilon \rightarrow 2 \pi$ gives a vanishing GMPFT which is plausible since the whole domain is absorbing the Brownian particle. The Eq.~(\ref{expAnnulusGMFPT}) is shown in Fig.~\ref{fig_annulus_1d}b and compared with the numerical solution for $\delta= 10^{-3}$ and several values for the escape angle $\varepsilon \ge 10^{-2}$. We conclude that the prediction of the pseudo-one-dimensional approximation is valid in the limit $\delta \ll 1$ for a fixed value of escape angle $\varepsilon$.

The NE expression of the GMFPT given by Eq.~(\ref{GMFPTAnnulus}) is then valid for small values of $\varepsilon$ such that $-\delta \ln \varepsilon > 1$ whereas the pseudo-1d expression of the GMFPT given by Eq.~(\ref{expAnnulusGMFPT}) is valid for small values of $\delta$ such that $-\delta \ln \varepsilon \ll 1$. These two validity domains are shown in Fig.~\ref{fig_annulus_minimum} and a comparison between analytical solutions and numerical solution in both domains is proposed in Fig.~\ref{fig_annulus_1d}a. We can then see that the minimum value of the GMFPT is reached in the narrow escape regime.

In the $\Delta \ll {\cal R}$ limit, the maximal MFPT is reached at $\theta = \pi$ and its expression is deduced from Eq.~(\ref{expAnnulusFPTSurf})
\begin{equation}
\label{expAnnulusMMFPT}
\frac{D_\Delta t_{\rm max}}{{\cal R}^2} = \frac{D_\Delta t(\pi)}{{\cal R}^2}  \underset{\Delta \ll {\cal R}}{\simeq} \frac{(2\pi-\varepsilon)^2}{8}.
\end{equation}
A numerical verification of this expression is also shown in Fig.~\ref{fig_annulus_1d}b for $\delta= 10^{-3}$ and several values of escape angle $\varepsilon \ge 10^{-2}$. In the $\varepsilon \rightarrow 0$ and $\delta \rightarrow 0$ limits, such that $- \delta \ln \varepsilon \le 1$, this expression is equivalent to Eq.~(\ref{MMFPTAnnulus00}). The NE expression of the MMFPT given by Eq.~(\ref{MMFPTAnnulus}) is then valid for small values of $\varepsilon$ such that $-\delta \ln \varepsilon > 1$ whereas the pseudo-1d expression of the GMFPT given by Eq.~(\ref{expAnnulusMMFPT}) is valid for small values of $\delta$ such that $-\delta \ln \varepsilon \ll 1$. These validity domains are shown in Fig.~\ref{fig_annulus_1d}b.

%%% SECTION 3

\section{Narrow escape problem for the two-shell geometry}
\label{sec3}

In this section, we investigate the NE problem for the two-shell geometry where the search domain is separated into two concentric regions: the outer shell of width $\Delta$ for a radius in ${\cal R}-\Delta \le |{\bf x}| \le {\cal R}$ denoted as $\Omega_\Delta$ and the inner shell for a radius $|{\bf x}| \le {\cal R}-\Delta$. The inner boundary located at $|{\bf x}|={\cal R}-\Delta$ is denoted as $\partial \Omega_\Delta$ whereas the outer boundary located at $|{\bf x}|={\cal R}$ is denoted as $\partial \Omega$. The escape region is located on the outer boundary for $|\theta| \le \varepsilon/2$ and is denoted as $\partial \Omega_\varepsilon$. The diffusivity of the Brownian particle is different in the two shells, denoted as $D_0$ and $D_\Delta$ in the inner shell and the outer shell respectively (see Fig.\ref{fig_domains}c). The MFPT for a diffusive particle starting in the outer shell and in the inner shell is denoted as $t_1({\bf x})$ and $t_2({\bf x})$ respectively. The equations satisfied by the functions $t_i({\bf x})$ are deduced from the Eq.~(\ref{linkMFPTprob}) and the backward equations for the probability density function~\cite{redner_2001_guide}
\begin{gather}
D_\Delta \nabla^2 t_1({\bf x}) = -1, \quad {\bf x} \in \Omega_\Delta \label{FPTTL1} \\
D_0 \nabla^2 t_2({\bf x}) = -1, \quad {\bf x} \in \Omega \backslash \Omega_\Delta  \label{FPTTL2} \\
t_1({\bf x}) =  t_2({\bf x}) \quad {\rm and} \quad D_\Delta {\bf n} \cdot \nabla t_1({\bf x}) =  D_0 {\bf n} \cdot \nabla t_2({\bf x}) , \quad {\bf x} \in \partial \Omega_\Delta \label{FPTTL3}\\
t_1({\bf x}) = 0, \quad {\bf x} \in \partial \Omega_\varepsilon \quad {\rm and} \quad {\bf n} \cdot \nabla t_1({\bf x}) = 0, \quad {\bf x} \in \partial \Omega \backslash \partial \Omega_\varepsilon. \label{FPTTL4}
\end{gather}
The boundary conditions (\ref{FPTTL3}) on the inner boundary guarantee the continuity of the concentration and the flux of the Brownian particles across the interface, implying the continuity of the probability density function $p({\bf y},\tau|{\bf x})$ and the normal component of the current defined by the Fick's law as $-D({\bf x}) \nabla p({\bf y},\tau|{\bf x})$. Note that the model studied here and the resulting equations (\ref{FPTTL1}-\ref{FPTTL4}) are different from the diffusion equation with a spatially varying diffusion constant $D({\bf x})\nabla^2 t({\bf x})=-1$, which has been studied and solved exactly in Ref.~\cite{grebenkov_2016_universal}.

\begin{figure}[t]
\begin{center}
  \includegraphics[width=16cm]{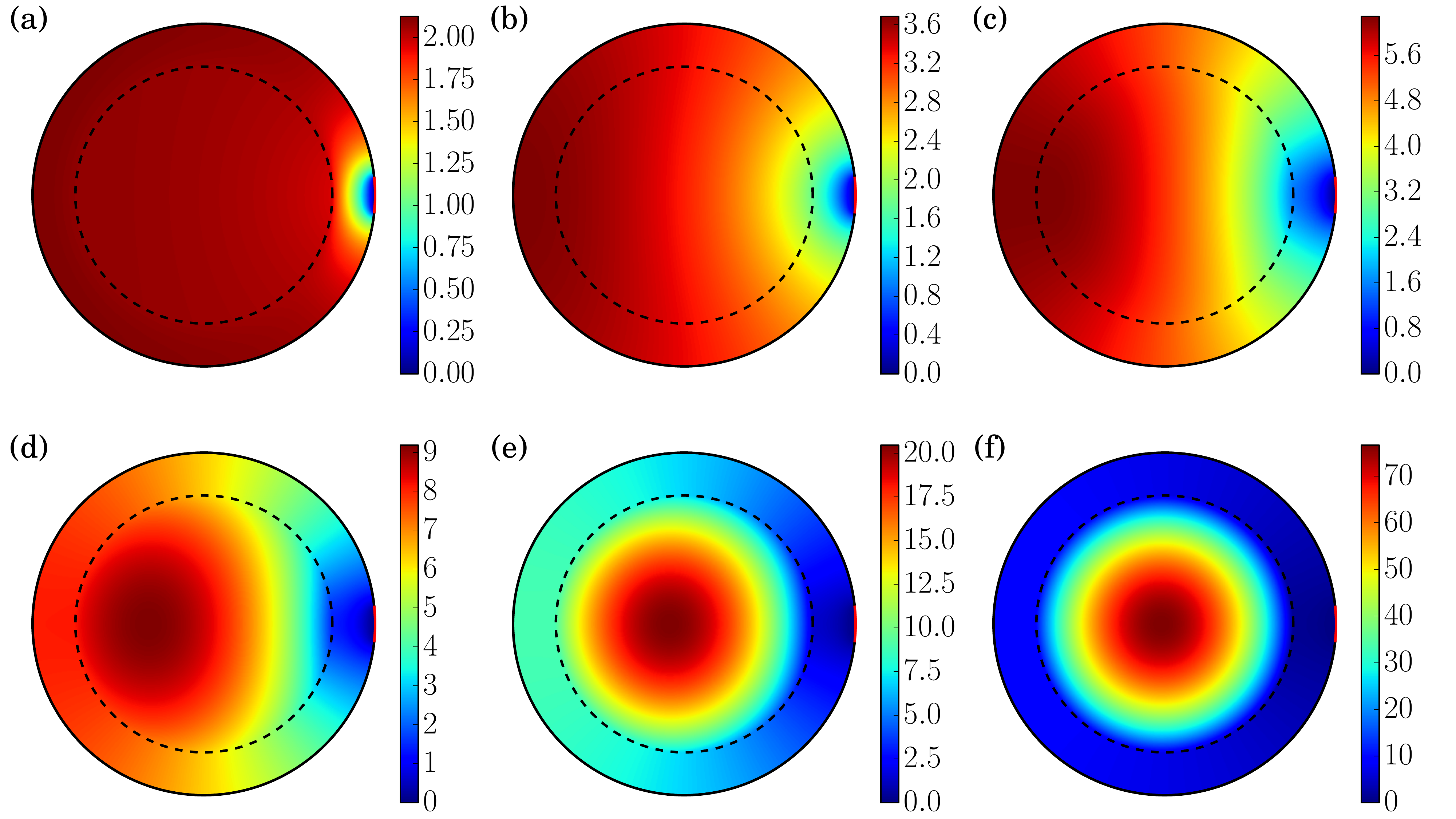}
  \caption{Numerical solution of the dimensionless mean first passage time (MFPT) for a diffusive particle starting at ${\bf x}$, $D_\Delta t({\bf x})/{\cal R}^2$, obtained from Eqs.~(\ref{FPTTL1}-\ref{FPTTL4}) with FreeFem++. This solution is plotted here for the width of outer shell $\Delta=0.25 {\cal R}$ and for several outer shell diffusivities $D_\Delta$: {\bf (a)}~$D_\Delta=0.05 D_0$, {\bf (b)}~$D_\Delta= D_0$, {\bf (c)}~$D_\Delta=5 D_0$, {\bf (d)}~$D_\Delta=20 D_0$, {\bf (e)}~$D_\Delta=100 D_0$ and {\bf (f)}~$D_\Delta=500 D_0$. Note that the maximum value of the MFPT migrates from $(-{\cal R},0)$ to the center of the disk.\label{fig_twolayers_FPT}}
\end{center}
\end{figure}

Numerical solutions  of these PDEs are shown in Fig.~\ref{fig_twolayers_FPT} for $\Delta=0.25 {\cal R}$ and several values of the ratio $D_\Delta/D_0$. This problem has only three dimensionless parameters: the ratio of diffusivity between outer and inner shells $D_\Delta/D_0$, the width of the outer shell $\delta=\Delta/{\cal R}$ and the angle $\varepsilon$ of the escape region, after rescaling the lengths by the radius ${\cal R}$ and the time by ${\cal R}^2/D_\Delta$.

\subsection{The narrow escape limit ($\varepsilon \ll 1$)}

In the limit $\varepsilon\rightarrow 0$ the MFPT satisfies in sub-leading order Eq.~(\ref{expMFPT}) where the volume of the space is $|\Omega| = \pi {\cal R}^2$ and $D_0$ is replaced by the diffusivity of the region close to the hole: $D_\Delta$. The regular part of the Green's function $R({\bf x}|{\bf x_0})$ is calculated in the appendix \ref{appendixTwoShells} and its expression is given by Eqs.~(\ref{Rtwoshells1}-\ref{Rtwoshells2}) in terms of (dimensionless) polar coordinates. Its value at the center of the escape region is then equal to
\begin{equation}
\label{RGreenTL}
R({\bf x_0}|{\bf x_0})= \frac{1}{8\pi} + \frac{(D_\Delta-D_0)(1-\delta)^4}{8\pi D_0} + \frac{2}{\pi}\sum_{n=1}^\infty \frac{(D_\Delta-D_0)(1-\delta)^{2n}}{n\left\{2D_0 + (D_\Delta-D_0)\left[1-(1-\delta)^{2n}\right]\right\}}.
\end{equation}
We observe that the value $R({\bf x_0}|{\bf x_0})$ goes to $1/8\pi$ when $D_\Delta \rightarrow D_0$ or when $\Delta \rightarrow {\cal R}$. In fact, in these two limits, the physical problem is identical to the disk geometry discussed in Sec.~\ref{sec1}. In the limit $D_\Delta \gg D_0$, we note that $R({\bf x_0}|{\bf x_0})$ diverges, implying a divergence of the GMFPT. This is due to the fact that the particles starting at $|{\bf x}|<{\cal R}-\Delta$ are slowed down by the small diffusivity in the central core. Moreover, we note that in the limit $\Delta \rightarrow 0$, the value of $R({\bf x_0}|{\bf x_0})$ diverges as $\ln (\Delta/{\cal R})$. The expression of the GMFPT is then given by Eq.~(\ref{expGMFPT})
\begin{equation}
\label{GMFPT2shells}
\frac{D_\Delta \langle t\rangle}{{\cal R}^2} \underset{\varepsilon \ll 1}{\simeq} - \ln \frac{\varepsilon}{4} + \frac{1}{8} + \frac{D_\Delta-D_0}{8D_0}(1-\delta)^4 + 2 \sum_{n=1}^\infty \frac{(D_\Delta-D_0)(1-\delta)^{2n}}{n\left\{2D_0 + (D_\Delta-D_0)\left[1-(1-\delta)^{2n}\right]\right\}}.
\end{equation}
From the Eqs.~(\ref{expMFPT}) and~(\ref{Rtwoshells2}), the expression of the CMFPT for a diffusive particle starting at the center of the disk is in the narrow escape limit
\begin{equation}
\label{CMFPT2shells}
\frac{D_\Delta t({\bf 0})}{{\cal R}^2} \underset{\varepsilon \ll 1}{\simeq} - \ln \frac{\varepsilon}{4} + \frac{1}{4} + \frac{D_\Delta-D_0}{4D_0}(1-\delta)^2 + 2 \sum_{n=1}^\infty \frac{(D_\Delta-D_0)(1-\delta)^{2n}}{n\left\{2D_0 + (D_\Delta-D_0)\left[1-(1-\delta)^{2n}\right]\right\}}.
\end{equation}

\begin{figure}[t]
\begin{center}
  \includegraphics[width=16cm]{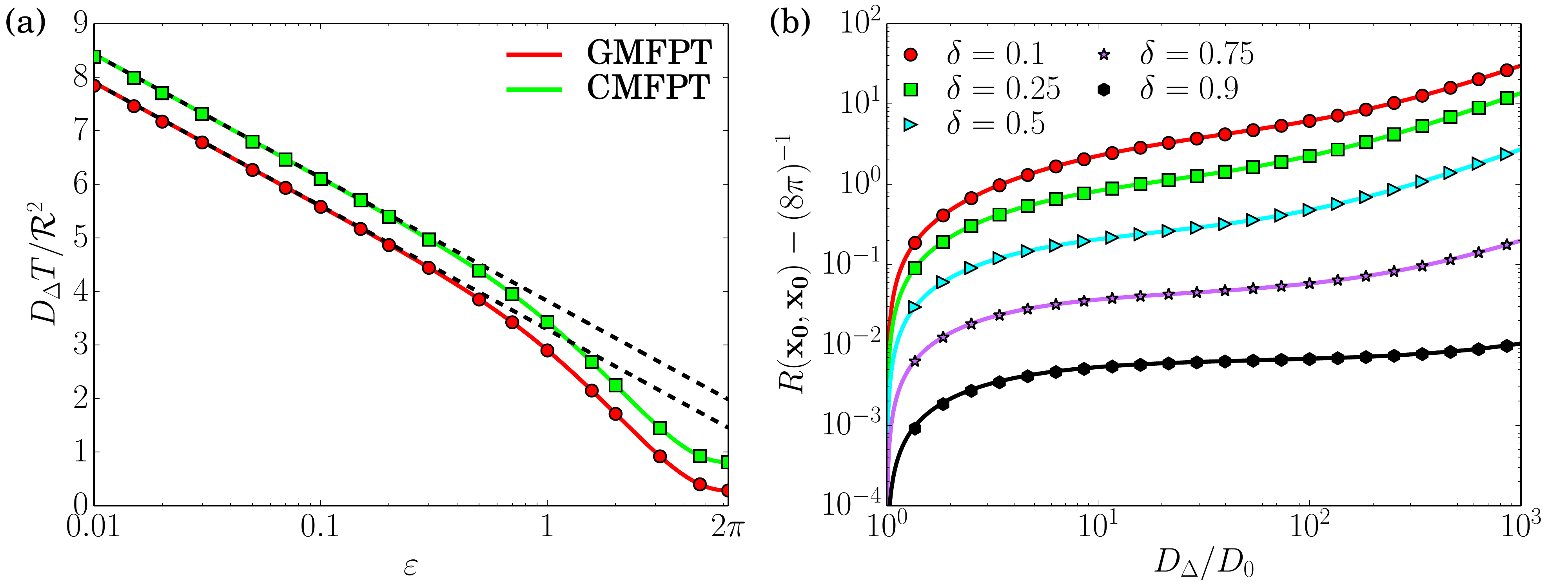}
  \caption{{\bf (a)}~Numerical solution of the dimensionless global mean first passage time (GMFPT) $D_\Delta \langle t \rangle/{\cal R}^2$ (circles) and the dimensionless mean first passage time starting at the center (CMFPT) $D_\Delta t({\bf 0})/{\cal R}^2$ (squares) obtained by the numerical integration of the Langevin equation (see appendix \ref{appendixLangevin}) as a function of $\varepsilon$. The parameters took for this figure are $\Delta=0.25 {\cal R}$ and $D_\Delta/D_0 = 5$. Full lines represent the numerical solutions obtained with the PDE solver and dashed lines represent the narrow escape solutions Eqs.~(\ref{GMFPT2shells}) and~(\ref{CMFPT2shells}) for the GMFPT and the CMFPT respectively. All three solutions are compatible in the $\varepsilon \ll 1$ limit, which validates (i) the 2d algorithm for diffusion in discontinuous media, (ii) the numerical solution of the PDEs by FreeFem++ and (iii) the validity of the narrow escape limit for the $\varepsilon \ll 1$ expansion. {\bf (b)}~$R({\bf x_0}|{\bf x_0})-1/8\pi$ as a function of $D_\Delta/D_0 \ge 1$ for several values of $\delta=\Delta/{\cal R}$. Full lines represent the analytical prediction from Eq.~(\ref{RGreenTL}) and symbols represent the numerical estimates obtained with fits to the results produced by the PDE solver. \label{fig_twolayers_NE}}
\end{center}
\end{figure}

The result for the GMFPT and CMFPT, Eqs.~(\ref{GMFPT2shells}) and (\ref{CMFPT2shells}), valid in the narrow escape limit $\varepsilon\rightarrow 0$, are shown in Fig.~\ref{fig_twolayers_NE}a as a function of $\varepsilon$ for $D_\Delta = 5D_0$ and $\delta=0.25$. The analytical solutions are compared with numerical solutions of the PDEs (\ref{FPTTL1}-\ref{FPTTL4}) obtained with FreeFem++ and with the results of Brownian particle simulations. In figure~\ref{fig_twolayers_NE}b, the analytical predictions for $R({\bf x_0}|{\bf x_0})$ given by Eq.~(\ref{RGreenTL}) are compared with the numerical fits of the constant sub-leading order of the GMFPT obtained with FreeFem++.

For the numerical simulation of Brownian motion, we apply the algorithm presented in the appendix \ref{appendixLangevin} to the radial coordinate of the Brownian particle. When the particle is on the inner boundary defined by the identity $|{\bf x}|={\cal R}-\Delta$, the probability to go to the inner or the outer shell is given respectively by Eq.~(\ref{pleft}) and Eq.~(\ref{pright}). The movement is then decomposed into radial and orthoradial coordinates assuming that - for very small displacement - the inner boundary is approximatively a wall with a zero curvature. This algorithm is valid for any values of $\delta$ and $\varepsilon$. The relative error of these stochastic simulations is inversely proportional to the square-root of their number of realizations $N=10^6$. The relative error is then of order $10^{-3}$.

Furthermore, when $\varepsilon=2\pi$, the mixed Neumann-Dirichlet boundary condition becomes a fully absorbing (Dirichlet) boundary condition implying that the MFPT function is radial. The GMFPT and CMFPT expressions are respectively
\begin{gather}
\label{GMFPT2shellsFull}
\frac{D_\Delta \langle t \rangle}{{\cal R}^2} \underset{\varepsilon = 2 \pi}{=}  \frac{1}{8} + \frac{D_\Delta-D_0}{8D_0} (1-\delta)^4,\\
\label{CMFPT2shellsFull}
\frac{D_\Delta t({\bf 0})}{{\cal R}^2} \underset{\varepsilon = 2 \pi}{=}  \frac{1}{4} + \frac{D_\Delta-D_0}{4D_0} (1-\delta)^2.
\end{gather}
The GMFPT and the CMFPT expressions in the narrow escape limit given respectively by Eqs.~(\ref{GMFPT2shells}) and~(\ref{CMFPT2shells}) can be decomposed into two terms: respectively the GMFPT and the CMFPT to reach the surface given by Eqs.~(\ref{GMFPT2shellsFull}) and~(\ref{CMFPT2shellsFull}) and the GMFPT to reach the escape region starting on the surface, which satisfies the expression
\begin{equation}
\frac{D_\Delta}{{\cal R}^2}t(\partial \Omega \rightarrow \partial \Omega_\varepsilon) \simeq - \ln \frac{\varepsilon}{4} + 2 \sum_{n=1}^\infty \frac{(D_\Delta-D_0)(1-\delta)^{2n}}{n\left\{2D_0 + (D_\Delta-D_0)\left[1-(1-\delta)^{2n}\right]\right\}}.
\end{equation}

Moreover, the starting point that has the maximum MFPT is not always located at $-{\bf x_0}=(-{\cal R},0)$ [see Fig.~\ref{fig_twolayers_FPT}]. In fact, if the diffusivity of the inner shell is much bigger than the one of the outer shell ($D_\Delta>D_0$), the time spent in the inner shell can be bigger than the one spent in the outer shell which implies that the MFPT to reach the escape region - in the outer shell - is maximized in the inner shell. Because of the symmetries of the problem (concentric and symmetric with respect to the horizontal axis), the maximum must be located on the horizontal axis, since it depends on the distance to the escape region, and more precisely on the segment between $-{\bf x_0}$ and ${\bf 0}$. We denote the distance between the starting point and the center of the disk that gives the maximum MFPT (MMFPT) with $r_{\rm max}$ and with $D_\Delta^c>D_0$ the lower limit of the diffusivity $D_\Delta$ for which the starting point with the MMFPT is located in the inner shell. If $D_\Delta < D_\Delta^c$, the MMFPT starting point is located in the outer shell ($|{\bf x}| > {\cal R} - \Delta$) by definition of $D_\Delta^c$. The value of $r_{\rm max}$ is then restricted between ${\cal R}$ and ${\cal R}-\Delta$. Since the MFPT is an increasing function of the distance to the escape region, when the diffusivity is constant, the MMFPT is then necessarily situated at $-{\bf x_0}$ and $r_{\rm max} = {\cal R}$ [see Fig.~\ref{fig_twolayers_FPT}c]. If $D_\Delta > D_\Delta^c$, the MMFPT starting point is located in the inner shell ($|{\bf x}| < {\cal R} - \Delta$) by definition of $D_\Delta^c$. The value of $r_{\rm max}$ is then lesser than ${\cal R} - \Delta$. For a diffusivity $D_\Delta \gtrsim D_\Delta^c$, the MMFPT starting point is located in the maximum distance to the escape region within the inner shell implying that $r_{\rm max} \lesssim {\cal R} - \Delta$ [see Fig.~\ref{fig_twolayers_FPT}d].

When $D_\Delta \gg D_\Delta^c$, the MFPT can be decomposed into the sum of the MFPT to reach the inner boundary ${\bf x} = {\cal R} - \Delta$ and the MFPT to reach the escape region from this last point. The MMFPT starting point is then located at the center of the disk which implies that $r_{\rm max} = 0$ [see Fig.~\ref{fig_twolayers_FPT}f]. Since the MMFPT starting point stays for $D_\Delta > D_\Delta^c$ in a region of constant diffusivity, $r_{\rm max}$ decreases continuously from ${\cal R} - \Delta$ to $0$.

\begin{figure}[t]
\begin{center}
  \includegraphics[width=16cm]{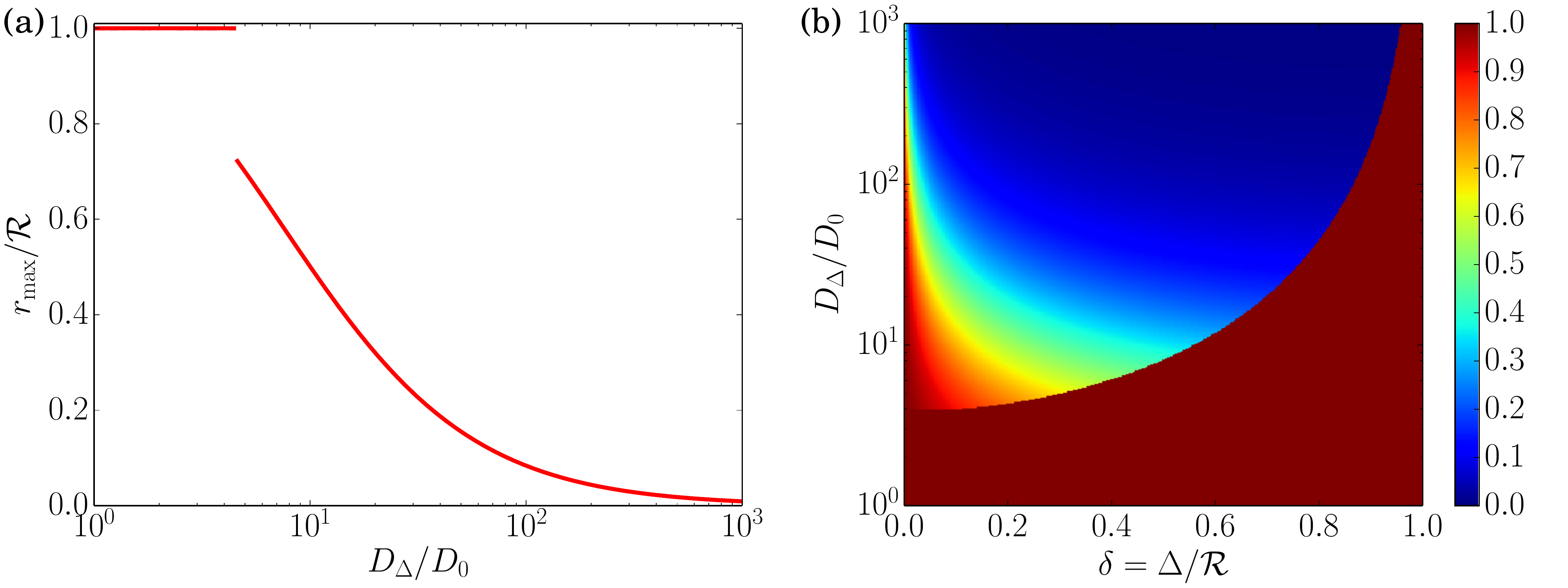}
  \caption{{\bf (a)}~Dimensionless distance of the maximal mean first passage time (MMFPT) to the center $r_{\rm max}/{\cal R}$ for $\Delta=0.25 {\cal R}$ and $\varepsilon=0.05$. For $D_\Delta<D_\Delta^c$, the MMFPT is located at the point $(-{\cal R},0)$ for which $r_{\rm max}={\cal R}$. At $D_\Delta=D_\Delta^c$, the position of the MMFPT migrates discontinuously to the inner shell and moves continuously towards the center of the disk in the limit $D_\Delta \gg D_0$. {\bf (b)}~Dimensionless distance of the MMFPT to the center $r_{\rm max}/{\cal R}$ as a function of the width $\Delta$ and the ratio $D_\Delta$. The MMFPT is always located at $(-{\cal R},0)$ in the limits $D_\Delta \rightarrow D_0$ and $\Delta \rightarrow {\cal R}$ corresponding to the disk geometry. This discontinuity is more pronounced for increasing values of $\Delta$. The $D_\Delta^c$ value seems to be a continuous function of $\Delta$. \label{fig_maximum}}
\end{center}
\end{figure}

To summarize, the location of the MMFPT moves discontinuously from the position $-{\bf x_0}$ ($r_{\rm max}={\cal R}$) for $D_\Delta < D_\Delta^c$ to a position inside the inner shell ($r_{\rm max} < {\cal R}-\Delta$) for $D_\Delta > D_\Delta^c$ and approaches continuously the center of the disk ($r_{\rm max} =0$) in the limit $D_\Delta \gg D_\Delta^c$. This behaviour is demonstrated in Fig.~\ref{fig_maximum}a for $\Delta = 0.25 {\cal R}$ and $\varepsilon = 0.05$. The limit value of the diffusivity at the discontinuity is equal to $D_\Delta^c \simeq 4.5 D_0$. In Fig.~\ref{fig_maximum}b, we show the position of the maximum $r_{\rm max}$ with respect to the parameters $\Delta/{\cal R}$ and $D_\Delta$ for the escape angle $\varepsilon = 0.05$. The function $D_\Delta^c(\Delta)$ can be extracted from the discontinuity of the location of the MMFPT and appears to be continuous. The discontinuity is less pronounced for $\Delta \ll {\cal R}$ since the width of the outer shell is small while the discontinuity is more pronounced for $\Delta \simeq {\cal R}$ but occurring for a larger diffusivity - for a geometry close to the disk.

\subsection{The thin outer shell limit ($\Delta \ll {\cal R}$)}

\begin{figure}[t]
\begin{center}
  \includegraphics[width=16cm]{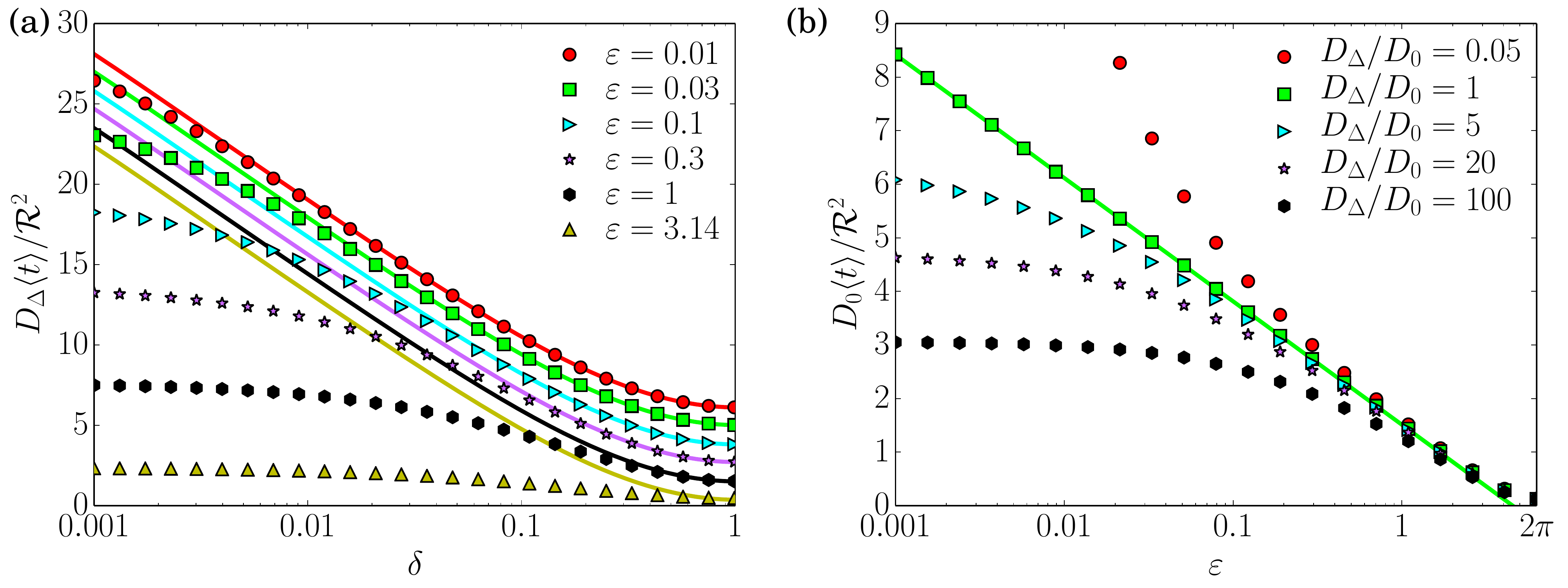}
  \caption{{\bf (a)}~Numerical solution (symbols) of the dimensionless global mean first passage time (GMFPT) $D_\Delta \langle t \rangle/{\cal R}^2$ as a function of $\delta=\Delta/{\cal R}$ for several values of $\varepsilon$ and $D_\Delta/D_0 = 5$. It is compared to the narrow escape solution (straight lines) satisfying the Eq.~(\ref{GMFPT2shells}) valid if and only if $\Delta$ is not too small. In the $\Delta \ll {\cal R}$ limit, the GMFPT converges through a constant value. {\bf (b)}~Numerical value (symbols) of the GMFPT for $\delta =10^{-3}$ plotted with respect to $\varepsilon$ for several ratios $D_\Delta/D_0$. For $D_\Delta > D_0$, the GMPT has a constant limit for $\varepsilon \rightarrow 0$ whereas the narrow escape solution (\ref{circularGMFPT}) is valid for $D_\Delta = D_0$. Moreover, the GMFPT diverges in the limit $\varepsilon \rightarrow 0$ for $D_\Delta < D_0$. \label{fig_twolayers_SM}}
\end{center}
\end{figure}

Next we calculated the GMFPT numerically with FreeFEM++ and show the result in Fig.~\ref{fig_twolayers_SM}a as a function of the width $\Delta$ for fixed values of the escape angle $\varepsilon$ and for $D_\Delta=5D_0$. We compare it with the narrow escape expression (\ref{GMFPT2shells}) and note that it is valid only for not too small values of $\Delta$ since the analytical solution diverges when $\Delta \ll {\cal R}$ while the numerical solution approaches a constant. This behavior shows that the diffusive motion is enhanced by the excursions in the thin outer shell ($\Delta \ll {\cal R}$), which implies a finite GMPFT to reach a small escape region on the outer boundary. This kind of behavior was already observed for surface-mediated diffusion problems \cite{benichou_2010_optimal} where the leading order of the narrow escape problem is constant and no more logarithmic divergent with the escape angle $\varepsilon$. This behavior only happens for $D_\Delta > D_0$ as shown in Fig.~\ref{fig_twolayers_SM}b. In fact, for $D_\Delta=D_0$ the two-shell geometry reduces to the disk geometry studied in section \ref{sec1} presenting the classical logarithmic behavior. And for $D_\Delta<D_0$, the behavior is super-logarithmic since the GMFPT diverges much faster than the logarithmic behavior of the $D_\Delta=D_0$ case: the Brownian motion in the outer shell is slowed down due to the diffusion constant value, smaller than in the inner shell, and the MFPT to reach the escape region located in it, is strongly increased.

\begin{figure}[t]
\begin{center}
  \includegraphics[width=16cm]{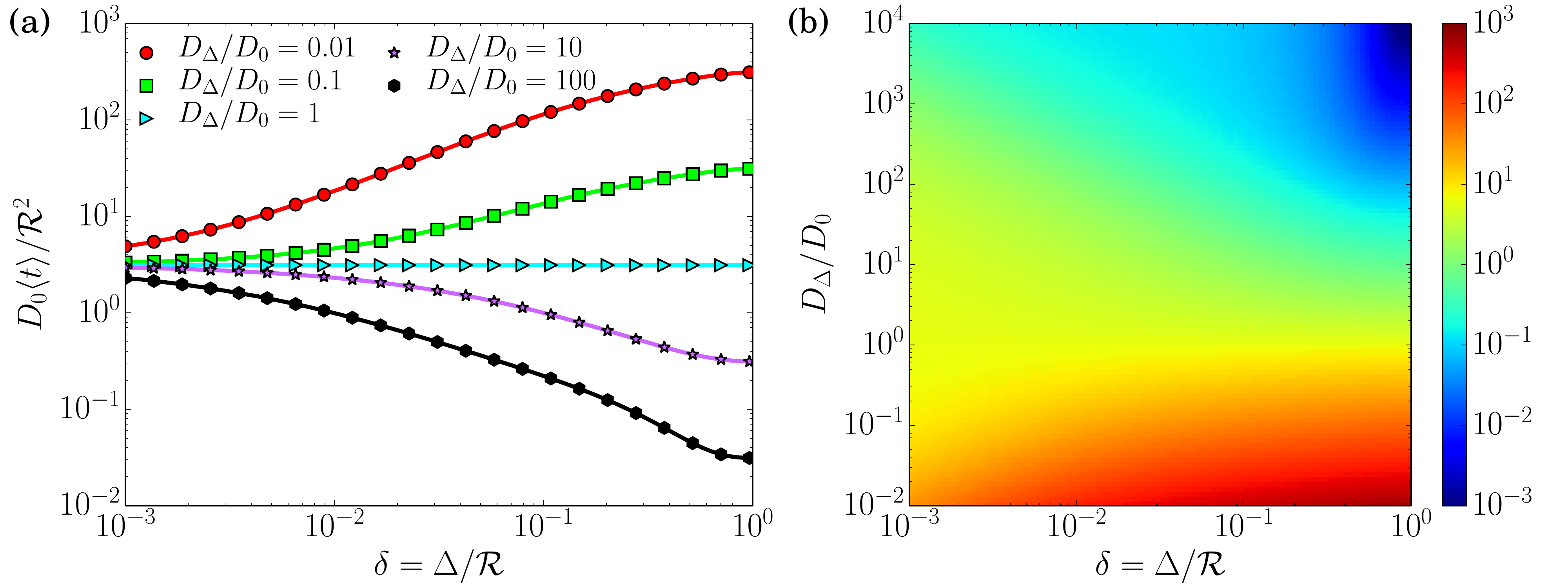}
  \caption{{\bf (a)}~Numerical solution of the dimensionless global mean first passage time (GMFPT) $D_0 \langle t \rangle/{\cal R}^2$ as a function of $\delta = \Delta/{\cal R}$ for several values of $D_\Delta/D_0$ and $\varepsilon = 0.2$. This function is monotonous and strictly decreasing if and only if $D_\Delta>D_0$. The same behavior can be seen for the mean first passage time for a particle starting at the center (CMFPT) $D_0 t({\bf 0})/{\cal R}^2$ (not shown here). {\bf (b)}~Numerical solution of the dimensionless GMFPT as a function of $\delta = \Delta/{\cal R}$ and $D_\Delta/D_0$ for $\varepsilon=0.01$. This two-variable function is strictly monotonous with respect to both parameters. \label{fig_twolayers_GMFPT}}
\end{center}
\end{figure}

Moreover, Fig.~\ref{fig_twolayers_GMFPT} shows that the GMFPT is a monotonic function of $\Delta/{\cal R}$ and $D_\Delta/D_0$ for a fixed value of $\varepsilon$. The same observation holds for the CMFPT (the data is not shown here). The GMFPT as a function of $\Delta$ is a strictly decreasing function if and only if $D_\Delta>D_0$. When the value of $\Delta$ is increased, the GMFPT is decreased if $D_\Delta>D_0$ since the area with the biggest diffusivity (the outer shell) increases. Strictly the opposite behavior happens when $D_\Delta<D_0$ since the area with the lowest diffusivity increases. The GMFPT as a function of $D_\Delta/D_0$ is a decreasing function since the MFPT is inversely proportional to the diffusivity. Then, there is no minimum of the MFPT for the two-shell geometry, {\it i. e.} for $\Delta\ne 0$ and $\Delta\ne {\cal R}$.

The limiting case $D_0 \ll D_\Delta$ is not equivalent to the annulus geometry studied in the section \ref{sec2} since the probability to find a particle in the inner shell is not zero. In fact, the diffusive particle can enter in the inner shell and the stationary probability (without the hole) to find it in the inner shell is the same as in the outer shell and equal to~$1/|\Omega|$, at the contrary of the annulus geometry for which the Brownian particle is excluded from the inner shell. The MFPT starting far from the escape region in the outer shell will then be impacted by this difference: the particle will have a high probability, equal to $(1-\delta)^2$ from ergodicity, to enter in the inner shell which plays the role of a trap where the particle stays a long time since $D_0 \ll D_\Delta$. Since the inner boundary does not play the role of a barrier but just separates two regions with different spreading of particles, the MFPT obtain for the annulus geometry will be considerably modified and then explain the differences between both solutions.

\section{Discussion}
\label{sec4}

We have studied the narrow escape problem for two different disk-like geometries with an inner and an outer region, the latter of width $\Delta$. The annulus geometry for which the Brownian particle is excluded from the center shell and the two-shell geometry where the particle has two different diffusion constants. The narrow escape problem consists to look at the asymptotic behavior of the MFPT when the escape region is reduced to a small escape angle $\varepsilon \ll 1$ when the width $\Delta$ is fixed. Furthermore, we analyzed the behavior of the MFPT in the thin outer shell limit $\Delta \ll {\cal R}$ when the size of the escape region $\varepsilon$ is fixed.
 
For the annulus geometry, both limits are compatible such that the expression of the global mean first passage time (GMFPT) is given at the leading order by
\begin{equation}
\frac{D_\Delta \langle t \rangle}{{\cal R}^2} = -2 \delta \ln \varepsilon + \frac{\pi^2}{3} + {\cal O}(\delta,\varepsilon),
\end{equation}
for the dimensionless width $\delta= \Delta / {\cal R}$. Whereas the GMFPT is an increasing function of $\varepsilon$, it can be optimized with respect to the width $\Delta$ of the annulus. The minimum of GMFPT can be explained by the reduction of the exploration area for Brownian particles starting close to the escape region while the particles starting far from it remain {\it jammed} for a longer time when the obstacle is large.

For the two-shell geometry, both limits are not interchangeable since the GMFPT is equal to $\langle t \rangle \propto -\ln(\varepsilon/4) +C(\Delta)$ in the narrow escape limit with a sub-leading order term that diverges in the thin outer shell limit: $C(\Delta) \underset{\Delta \rightarrow 0}{\propto} \ln (\Delta/{\cal R})$, whereas the limit of the GMFPT is constant only when $D_\Delta>D_0$. A similar behavior was recently derived \cite{grebenkov_2017_effects} for the MFPT to reach a small target of height $\varepsilon$ on a cylinder of radius $\Delta$, exhibiting similar asymptotically logarithmic features. Moreover, we found that the distance of the starting point with the maximum MFPT jumps discontinuously from ${\cal R}$, corresponding to a point opposite to the escape region, to a value close to ${\cal R}-\Delta$, corresponding to a point within the inner shell, when the outer diffusion constant $D_\Delta$ is increased from $D_0$ to a sufficiently high value.

It turns out that the GMFPT is a strictly monotonous function for all three main parameters : $\varepsilon$, $\Delta/{\cal R}$ and $D_\Delta/D_0$. This is in contrast to the optimizability of the MFPT reported in Refs.~\cite{schwarz_2016_optimality, schwarz_2016_numerical}. In a model for spatially inhomogeneous intermittent search that alternates stochastically between ballistic and diffusive motion where ballistic motion is only radial in the inner shell and multi-directional in the outer shell, it was found that the MFPT for the narrow escape problem can be minimized for a thin outer shell, provided the outward radial transport in the inner shell is enhanced. Thus in the model considered in Refs.~\cite{schwarz_2016_optimality, schwarz_2016_numerical}, it is the enforced re-injection of the searcher into the outer shell that is crucial for the optimization effect. It leads to an increased probability to find the particle in the outer shell, which is where the target is: the narrow escape region. On the other hand, the optimizability of the MFPT found for surface mediated diffusion~\cite{benichou_2010_optimal} is due to a low desorption and a high absorption probabilities to the surface also enforcing the particle to stay on the surface, where again the target is. Analogously, we mention the result of Ref.~\cite{grebenkov_2017_diffusive} concerning the optimization of the MFPT with respect to the range of attractive and repulsive potential on the surface which creates a competition between bulk and boundary events.

This analysis shows that it would be interesting to introduce for the model considered in this paper a similar mechanism leading to an increased probability to find the target in the outer shell. A straightforward possibility would be to make the boundary between the inner and outer shell semi-permeable, allowing only transitions from the inner to the outer shell, to favor the finding of the target in the outer shell. A less constrained possibility would be to introduce appropriate asymmetric reflection and transmission probabilities (or energy barriers) for this boundary. Alternatively, one could consider stochastic resetting into the cortex region when leaving the outer shell~\cite{majumdar_2015_random}. Furthermore, these analytical solutions of the narrow escape problem for a two-shell geometry let suppose that the spatially inhomogeneous intermittent search~\cite{schwarz_2016_optimality, schwarz_2016_numerical} in this two-shell geometry could be solvable in the narrow escape limit using the same method. These extensions of the problem analyzed in the present paper will be considered in a forthcoming publication.

\section{Acknowledgment}

This work was performed with financial support from the Deutsche Forschungsgemeinschaft DFG within the Collaborative Research Center SFB 1027. M. M. thanks T. Gu\'erin for interesting discussions.

\appendix

\section{Narrow escape expression of the MFPT}
\label{appendixNE}

In this appendix, we present the derivation of the sub-leading term of the MFPT mainly inspired by the method of Ward and Keller~\cite{ward_1993_strong} which was improved by Pillay \etal \cite{pillay_2010_asymptotic} and Chevalier \etal \cite{chevalier_2011_firstpassage}. The escape region with an angle $\varepsilon \ll 1$ is considered as a perturbation on the boundary. The solution far from the hole (outer solution) takes on the form
\begin{equation}
\label{outer}
t({\bf x}) = -\frac{|\Omega|}{\pi D_0} \ln (\varepsilon/4) + \tau({\bf x}) +  {\cal O}(\varepsilon).
\end{equation}
where $\tau({\bf x})$ represents the space-dependent sub-leading term. The multiplicative constant $4$ is just taken for convenience. It is of order ${\cal O}(\varepsilon^0)$ usually included in the unknown $\tau$ function. From Eqs. (\ref{FPT1}-\ref{FPT3}), the function $\tau({\bf x})$ satisfies the equations
\begin{eqnarray}
D_0 \nabla^2 \tau({\bf x}) &= -1, &{\bf x} \in \Omega \label{outer1}\\
{\bf n} \cdot \nabla \tau({\bf x}) &= 0 , &{\bf x} \in \partial \Omega \label{outer2}.
\end{eqnarray}

The solution close to the hole (inner solution) is written in terms of the inner variable ${\bf \tilde x} = 2 ({\bf x} - {\bf x_0}) / \varepsilon {\cal R}$, such that the hole corresponds to the segment (independent of the value of $\varepsilon$) defined by $\tilde x = 0$ and $|\tilde y| \le 1$, and the reflecting boundaries corresponds to the half-lines defined by $\tilde x = 0$ and $|\tilde y| \ge 1$, considering $\varepsilon$ enough small to have a flat boundary. In these coordinates, one writes $t({\bf x}) \equiv v({\bf \tilde x})$. The equations for $v({\bf \tilde x})$ are then for $\varepsilon \rightarrow 0$
\begin{eqnarray}
\nabla^2 v({\bf \tilde x}) &= 0, &{\bf \tilde x} \in \Omega \\
v({\bf \tilde x}) &= 0, &{\bf \tilde x} \in \partial \Omega_\varepsilon \\
{\bf n} \cdot \nabla v({\bf \tilde x}) &= 0 , &{\bf \tilde x} \in \partial \Omega \backslash \partial \Omega_\varepsilon.
\end{eqnarray}
These equations can be solved in the elliptic coordinates ($\mu$, $\nu$) defined by $\tilde x = \sinh \mu \sin \nu $ and $\tilde y = \cosh \mu \cos \nu $. The Laplace's equation writes in this system of coordinates: $\partial_{\mu\mu} v + \partial_{\nu\nu} v = 0$. The escape region corresponds to $\mu = 0$ and $\nu \in [0,2\pi]$, implying the boundary condition $v(\mu = 0,\nu) =0$. The reflecting boundary corresponds to $\nu = 0$ and $\mu > 0$, involving the condition $\partial_\nu v(\mu,\nu=0) = 0$. The $\nu$-independent solution of this system is $v(\mu,\nu)= A \mu$. The outer and the inner solutions are matched in an intermediate region such that ${\bf x} \rightarrow {\bf x_0}$ and $|{\bf \tilde x}| \rightarrow \infty$ ({\it i. e.} $\mu \rightarrow \infty$ for elliptic coordinates). The inner solution behaves thus like
\begin{equation}
v({\bf \tilde x}) \simeq A \ln(2|{\bf \tilde x}|) = A \ln \frac{4|{\bf x} - {\bf x_0}|}{\varepsilon{\cal R}}.
\end{equation}
Comparing with the outer solution (\ref{outer}), this implies the value of $A = |\Omega|/(\pi D_0)$ and the origin of the factor $4$ is seen. The behavior of the function $\tau({\bf x})$ close to the hole is thus given by
\begin{equation}
\label{outer3}
\tau({\bf x} \rightarrow {\bf x_0}) = \frac{|\Omega|}{\pi D_0} \ln \frac{|{\bf x} - {\bf x_0}|}{{\cal R}}.
\end{equation}

Following Ref.~\cite{pillay_2010_asymptotic}, the pseudo Green's function $G({\bf x}|{\bf x_0})$ is introduced via the equations
\begin{gather}
\nabla^2 G({\bf x}|{\bf x_0}) = \frac{1}{|\Omega|} - \delta({\bf x} - {\bf x_0}), \quad {\bf x} \in \Omega \label{Green1app}\\
{\bf n} \cdot \nabla G({\bf x}|{\bf x_0}) = 0 , \quad {\bf x} \in \partial \Omega \label{Green2app}\\
G({\bf x} \rightarrow {\bf x_0} |{\bf x_0}) = -\frac{1}{\pi} \ln \frac{|{\bf x} - {\bf x_0}|}{{\cal R}} + R({\bf x_0}| {\bf x_0}), \label{Green3app}\\
\int_{\Omega} d{\bf x}\ G({\bf x}|{\bf x_0}) = 0 \label{Green4app},
\end{gather}
where $\delta({\bf x})$ represents the Dirac's distribution which can be omitted here since ${\bf x_0}$ belongs to the boundary of the domain and $R({\bf x_0}| {\bf x_0})$ is the (unknown) regular part of the Green's function at the center of the escape region. The function $\tau({\bf x})$ is then given by the relation obtained from Eqs.~(\ref{outer1}-\ref{outer2}) and~(\ref{outer3})
\begin{equation}
\label{defG}
\tau({\bf x}) = - \frac{|\Omega|}{D_0} G({\bf x}|{\bf x_0}) + \langle \tau \rangle,
\end{equation}
where the spatial average $\langle \tau \rangle$ satisfies
\begin{equation}
\langle \tau \rangle =  \frac{|\Omega|}{D_0} R({\bf x_0}|{\bf x_0}).
\end{equation}

The spatial average of the MFPT is then given by
\begin{equation}
\langle t \rangle =  \frac{|\Omega|}{\pi D_0} \left[ - \ln \frac{\varepsilon}{4} + \pi R({\bf x_0}|{\bf x_0}) \right],
\end{equation}
reproduced inside the main text via Eq.~(\ref{expGMFPT}) and the expression of the MFPT $t({\bf x})$ is
\begin{equation}
t({\bf x}) = \langle t \rangle - \frac{|\Omega|}{D_0} G({\bf x}|{\bf x_0}),
\end{equation}
corresponding to the given Eq.~(\ref{expMFPTlitt}) in the main text. These both equations depends on the pseudo Green's function defined by Eqs.~(\ref{Green1app}-\ref{Green4app}) reproduced in the main text via Eqs.~(\ref{Green1}-\ref{Green4}) where the Dirac distribution is removed.

\section{Computations of the regular part of the Green's function $R({\bf x}|{\bf x_0})$}

In this appendix, we will compute the expression of the regular part of the Green's function $R({\bf x}|{\bf x_0})$, defined by Eq.~(\ref{defofR}), for the three different geometries shown in Fig.~\ref{fig_domains}. We will start to a rewriting of Eqs.~(\ref{Green1}-\ref{Green4}) for this regular part. The Eq.~(\ref{Green3}) is satisfied by the definition of $R({\bf x}|{\bf x_0})$. Since the function $\ln |{\bf x} - {\bf x_0}|$ is a solution of the Laplace's equation, for all planar domains where ${\bf x} \ne {\bf x_0}$, the equation (\ref{Green1}) becomes
\begin{equation}
\label{RGreen1}
\nabla^2 R({\bf x}|{\bf x_0}) = \frac{1}{|\Omega|}, \quad {\bf x} \in \Omega.
\end{equation}
Moreover, the spatial average of $\ln(|{\bf x} - {\bf x_0}|/{\cal R})$ vanishes over the disk - via the integral over the orthoradial part - the equation (\ref{Green4}) gives
\begin{equation}
\label{RGreen3}
\int_{\Omega} d{\bf x}\ R({\bf x}|{\bf x_0}) = 0.
\end{equation}
The condition on the boundary of the domain (\ref{Green2}) is changed to
\begin{equation}
\label{RGreen2}
{\bf n} \cdot \nabla R({\bf x}|{\bf x_0}) = {\bf n} \cdot \nabla \frac{1}{\pi} \ln \frac{|{\bf x} - {\bf x_0}|}{{\cal R}}.
\end{equation}
We use the (dimensionless) polar coordinates $(r,\theta)$ defined as $r=|{\bf x}|/{\cal R}$ and ${\bf x} \cdot {\bf x_0}/{\cal R}^2 = r \cos \theta$. This definition allows us to rename the regular part of the Green's function as $R(r,\theta)$ without any loss of generality. Since the normal vector ${\bf n}$ is always radial for all studied geometries, the relation for the radial derivative of $\ln |{\bf x} - {\bf x_0}|$ at the point $(r,\theta)$ is then
\begin{equation}
\label{relLogSurf}
{\bf e_r} \cdot \nabla \frac{1}{\pi} \ln \frac{|{\bf x} - {\bf x_0}|}{{\cal R}} = \frac{1}{2\pi} \frac{2r - 2 \cos \theta}{r^2 - 2 r \cos \theta +1}.
\end{equation}

\subsection{Known solution: the disk geometry}
\label{appendixDisk}

The equation on the boundary of the domain (\ref{RGreen2}) rewrites
\begin{equation}
\frac{\partial R}{\partial r}(1,\theta) = \frac{1}{2\pi},
\end{equation}
which is $\theta$-independent. We can thus look for a radial solution, given by
\begin{equation}
\label{Rdisk}
R(r,\theta) = \frac{r^2}{4\pi} - \frac{1}{8\pi},
\end{equation}
involving the value at the center of the escape region $R({\bf x_0}|{\bf x_0}) \equiv R(1,0) = 1/8\pi$.

\subsection{The annulus geometry}
\label{appendixAnnulus}

The equations of the Green's function for the annulus geometry can be deduced from the Eqs.~(\ref{FPTannulus1}-\ref{FPTannulus4}) of the MFPT. Its regular part satisfies then the bulk equation (\ref{RGreen1}) and the {\it normalization condition} (\ref{RGreen3}). From Eq.~(\ref{RGreen2}), the boundary conditions respectively on the outer and the inner boundaries become
\begin{gather}
\frac{\partial R}{\partial r}(1,\theta) = \frac{1}{2\pi},\label{BCannulus1}\\
\frac{\partial R}{\partial r}(1-\delta,\theta)  = \frac{1}{2\pi} \frac{2(1-\cos \theta) - 2\delta}{2(1-\delta)(1-\cos\theta)+\delta^2}\label{BCannulus2},
\end{gather}
where we used the relation (\ref{relLogSurf}). Since this last equation is $\theta$-dependent, the solution $R(r,\theta)$ is not radially symmetric for this geometry. We thus look for a solution with the global form
\begin{equation}
R(r,\theta) = \sum_{n=0}^\infty f_n(r) \cos(n\theta)
\end{equation}
due to the periodicity of the polar coordinate $\theta$ and the symmetry of the problem under the $\theta \rightarrow - \theta$ transformation. The functions $f_n(r)$ should satisfy the equations resulting from the bulk equation (\ref{RGreen1})
\begin{gather}
f_n''(r) +\frac{1}{r} f_n'(r) - \frac{n^2}{r^2} f_n(r) =0, \quad n\ge1\\
f_0''(r) +\frac{1}{r} f_0'(r)  = \frac{1}{|\Omega|}.
\end{gather}
The global solution is thus given by $f_n(r) = a_n r^n + b_n r^{-n}$ (with $n\ge 1$) and $f_0(r) = r^2/(4|\Omega|) + a_0 + b_0 \ln r$, which just rewrites as
\begin{equation}
R(r,\theta) = \frac{r^2}{4|\Omega|} + a_0 + b_0 \ln r + \sum_{n=1}^\infty \left[a_n r^n + b_n r^{-n}\right]\cos(n\theta).
\end{equation}

We will now look at the expression for $a_n$ and $b_n$ coefficients. The first condition we use is given by the equation (\ref{RGreen3}). Since the average value of $\cos(n\theta)$ vanishes for $n\ge1$, the remaining equation is
\begin{equation}
\label{eq0annulus}
\int_{1-\delta}^1 dr\ r \left[\frac{r^2}{4|\Omega|} +a_0 + b_0 \ln  r\right] =0. 
\end{equation}
From the equation (\ref{BCannulus1}), we get the relation
\begin{equation}
\frac{1}{2|\Omega|} + b_0 + \sum_{n=1}^{\infty} (a_n-b_n) n \cos(n\theta) = \frac{1}{2\pi},
\end{equation}
which implies immediately from the orthogonality of $\cos(n\theta)$ that $a_n =b_n$ for $n\ge1$ and
\begin{equation}
\label{constb0annulus}
b_0 = - \frac{(1-\delta)^2}{2\pi\delta(2-\delta)}.
\end{equation}
This last relation implies the value of $a_0$ from the expression (\ref{eq0annulus})
\begin{equation}
a_0 = \frac{3}{8\pi} - \frac{\delta(2-\delta) + (1-\delta)^4\ln(1-\delta)}{2\pi\delta^2(2-\delta)^2}.
\end{equation}

The boundary condition at $r=1-\delta$ given by Eq.~(\ref{BCannulus2}) gives the relation, using the equality $a_n=b_n$,
\begin{equation}
\frac{1-\delta}{2|\Omega|} + \frac{b_0}{1-\delta} + \sum_{n=1}^{\infty} a_n n \left[(1-\delta)^{n-1} - (1-\delta)^{-n-1}\right] \cos(n\theta) = \frac{1}{2\pi} \frac{2(1-\cos \theta) - 2\delta}{2(1-\delta)(1-\cos\theta)+\delta^2}.
\end{equation}
We again use the orthogonality of $\cos(n\theta)$. For $n=0$, we get
\begin{equation}
\frac{1-\delta}{2|\Omega|} + \frac{b_0}{1-\delta} = \int_0^{2\pi} \frac{d\theta}{(2\pi)^2} \frac{2(1-\cos \theta) - 2\delta}{2(1-\delta)(1-\cos\theta)+\delta^2} = 0.
\end{equation}
This implies again that the Eq.~(\ref{constb0annulus}) is satisfied which ensures that a solution of the system exists. For $n\ne 0$, we get the relation
\begin{equation}
a_n n \pi \left[(1-\delta)^{n-1} - (1-\delta)^{-n-1}\right] = \int_0^{2\pi} \frac{d\theta}{2\pi} \frac{2(1-\cos \theta) - 2\delta}{2(1-\delta)(1-\cos\theta)+\delta^2} \cos(n\theta) = -(1-\delta)^{n-1},
\end{equation}
which implies the expression of $a_n$ as 
\begin{equation}
a_n = \frac{1}{n\pi\left[(1-\delta)^{-2n}-1\right]}.
\end{equation}

After determining all these constants, we get the general expression of the regular part of the Green's function $R(r,\theta)$
\begin{equation}
\label{Rannulus}
R(r,\theta) = \frac{r^2-2}{4\pi\delta(2-\delta)} + \frac{3}{8\pi} - \frac{(1-\delta)^4\ln(1-\delta)}{2\pi\delta^2(2-\delta)^2}-\frac{(1-\delta)^2 \ln r}{2\pi\delta(2-\delta)} + \frac{1}{\pi} \sum_{n=1}^{\infty} \frac{(r^n+r^{-n})(1-\delta)^{2n}}{n\left[1-(1-\delta)^{2n}\right]}\cos(n\theta).
\end{equation}
Its value at the center of the escape region ${\bf x_0}$ - $R({\bf x_0}|{\bf x_0}) \equiv R(1,0)$ - is then
\begin{equation}
R({\bf x_0}|{\bf x_0}) = \frac{3}{8\pi} - \frac{1}{4\pi \delta(2-\delta)} -\frac{(1-\delta)^4 \ln(1-\delta)}{2\pi\delta^2(2-\delta)^2} + \frac{2}{\pi} \sum_{n=1}^\infty \frac{(1-\delta)^{2n}}{n\left[1-(1-\delta)^{2n}\right]},
\end{equation}
which is reproduced in Eq.~(\ref{RGreenAnnulus}) of the main text.

In the limit $\delta \rightarrow 1$, the expression (\ref{Rannulus}) tends to the disk formula (\ref{Rdisk})
\begin{equation}
R(r,\theta) \underset{\delta \rightarrow 1}{\simeq} \frac{r^2}{4\pi} - \frac{1}{8\pi},
\end{equation}
for a radius which is let constant such that $r>1-\delta$. Moreover, the derivative of the function on the inner boundary (\ref{BCannulus2}) transforms as
\begin{equation}
\underset{\delta \rightarrow 1}{\lim} \frac{\partial R}{\partial r}(1-\delta,\theta)  = - \frac{1}{\pi} \cos \theta
\end{equation}
and is not applicable at the origin $r=0$, where $\theta$ is undefined. In the limit $\delta \rightarrow 1$, the regular part of the Green's function of the annulus geometry tends to the one of the disk geometry except at the origin point. This will not change the MFPT expression since it is a continuous function all over the domain.

\subsection{The two-shell geometry}
\label{appendixTwoShells}

The equations of the Green's function for the two-shell geometry can be deduced from the Eqs.~(\ref{FPTTL1}-\ref{FPTTL4}) of the MFPT. We define $R_1(r,\theta)$ and $R_2(r,\theta)$ the value of its regular part respectively on the inner shell and the outer shell in the (dimensionless) polar coordinates. The bulk equations verified by $R_i(r,\theta)$ are thus
\begin{gather}
\nabla^2 R_1(r,\theta) = \frac{1}{|\Omega|}, \quad r>1-\delta \label{RGreenTL1} \\
\nabla^2 R_2(r,\theta) = \frac{D_\Delta}{D_0 |\Omega|}, \quad r<1-\delta.  \label{RGreenTL2}
\end{gather}
The boundary conditions at $r=1-\delta$ are given by 
\begin{gather}
R_1(1-\delta,\theta) =  R_2(1-\delta,\theta) , \label{RGreenTL3} \\
D_\Delta \frac{\partial R_1}{\partial r}(1-\delta,\theta)  - D_0 \frac{\partial R_2}{\partial r}(1-\delta,\theta) = \frac{D_\Delta-D_0}{2 \pi} \frac{2(1-\cos\theta)-2\delta}{2(1-\delta)(1-\cos\theta)+\delta^2}, \label{RGreenTL4}
\end{gather}
where we used the relation (\ref{relLogSurf}), and the reflective boundary condition at $r=1$ becomes 
\begin{equation}
\frac{\partial R_1}{\partial r}(1,\theta) = \frac{1}{2\pi} \label{RGreenTL5}.
\end{equation}
Finally, the {\it normalization condition} gives the last relation, fixing the additive constant of $R$
\begin{equation}
\int_0^{2\pi} d\theta \left[ \int_0^{1-\delta} dr\ r R_2(r,\theta) + \int_{1-\delta}^1 dr\ r R_1(r,\theta) \right] = 0. \label{RGreenTL6}
\end{equation}

The general solutions of the bulk equations (\ref{RGreenTL1}) and (\ref{RGreenTL2}) can be written as
\begin{gather}
R_1(r,\theta) = \frac{r^2}{4\pi} + a_0 + b_0 \ln r + \sum_{n=1}^\infty \left[a_n r^n + b_n r^{-n}\right]\cos(n\theta),\\
R_2(r,\theta) = \frac{D_\Delta r^2}{4\pi D_0} + c_0 + d_0 \ln r + \sum_{n=1}^\infty \left[c_n r^n + d_n r^{-n}\right]\cos(n\theta).
\end{gather}
From the non singularity of the problem at $r=0$, we should impose $d_n =0$ for $n\ge0$ to have a non divergent solution $R_2(r,\theta)$. The boundary condition at $r=1$ given by Eq.~(\ref{RGreenTL5}) writes
\begin{equation}
\frac{1}{2\pi} + b_0 + \sum_{n=1}^\infty n \left(a_n - b_n\right)\cos(n\theta)  = \frac{1}{2\pi}.
\end{equation}
The orthogonality of $\cos(n\theta)$ imposes thus $b_0=0$ and $a_n=b_n$ for $n\ge1$. The continuity at $r=1-\delta$, expressed by the Eq.~(\ref{RGreenTL3}) becomes
\begin{equation}
\frac{(1-\delta)^2}{4\pi} + a_0 + \sum_{n=1}^\infty a_n \left[(1-\delta)^n + (1-\delta)^{-n}\right]\cos(n\theta) = \frac{D_\Delta (1-\delta)^2}{4\pi D_0} + c_0 + \sum_{n=1}^\infty c_n (1-\delta)^n \cos(n\theta).
\end{equation}
The orthogonality of $\cos(n\theta)$ imposes thus
\begin{gather}
a_0-c_0 = \frac{(D_\Delta-D_0)(1-\delta)^2}{4\pi D_0},\\
c_n=a_n \left[1+(1-\delta)^{-2n}\right].
\end{gather}
From the relation $\langle \cos(n\theta) \rangle = 0$, the condition (\ref{RGreenTL5}) becomes
\begin{equation}
\int_0^{1-\delta} dr\ r\left(\frac{D_\Delta r^2}{4\pi D_0} + c_0\right) + \int_{1-\delta}^1 dr\ r\left(\frac{r^2}{4\pi} + a_0\right) =0,
\end{equation}
which simplifies as
\begin{equation}
\frac{(D_\Delta-D_0)(1-\delta)^4+D_0}{8\pi D_0} - (a_0-c_0)(1-\delta)^2 + a_0=0.
\end{equation}
This equation gives thus
\begin{gather}
a_0 = \frac{(D_\Delta-D_0)(1-\delta)^4-D_0}{8\pi D_0},\\
c_0 = \frac{(D_\Delta-D_0)(1-\delta)^4-2(D_\Delta-D_0)(1-\delta)^2-D_0}{8\pi D_0}.
\end{gather}
Finally, to determine the $a_n$ and $c_n$ coefficients for $n\ge1$, we solve the condition (\ref{RGreenTL4}) at $r=1-\delta$
\begin{equation}
\begin{split}
D_\Delta \sum_{n=1}^\infty a_n n \left[ (1-\delta)^{n-1} - (1-\delta)^{-n-1} \right] \cos(n\theta) - D_0 \sum_{n=1}^\infty c_n n (1-\delta)^{n-1} \cos(n\theta)  =\\ \frac{D_\Delta-D_0}{2 \pi} \frac{2(1-\cos\theta)-2\delta}{2(1-\delta)(1-\cos\theta)+\delta^2}.
\end{split}
\end{equation}
The orthogonality of $\cos(n\theta)$ imposes thus
\begin{equation}
\begin{split}
a_n n \pi (1-\delta)^{n-1} \left\{D_\Delta \left[ 1 - (1-\delta)^{-2n}\right] - D_0  \left[1+(1-\delta)^{-2n}  \right] \right\}&= \\ \frac{D_\Delta-D_0}{2 \pi} \int_0^{2\pi} d\theta \ \frac{2(1-\cos\theta)-2\delta}{2(1-\delta)(1-\cos\theta)+\delta^2} \cos(n\theta) &= -(D_\Delta-D_0)(1-\delta)^{n-1},
\end{split}
\end{equation}
which implies the expressions of $c_n$ and $a_n$ coefficients
\begin{gather}
a_n = \frac{(D_\Delta-D_0)(1-\delta)^{2n}}{n \pi \left\{2D_0 + (D_\Delta-D_0)\left[1-(1-\delta)^{2n}\right]\right\}},\\
c_n = \frac{(D_\Delta-D_0)\left[(1-\delta)^{2n}+1\right]}{n \pi \left\{2D_0 + (D_\Delta-D_0)\left[1-(1-\delta)^{2n}\right]\right\}}.
\end{gather}

We have then determine the general expression of the regular part of the Green’s function $R(r,\theta)$ with
\begin{gather}
\label{Rtwoshells1}
R_1(r,\theta) = \frac{r^2}{4\pi} - \frac{1}{8\pi} + \frac{(D_\Delta-D_0)(1-\delta)^4}{8\pi D_0} + \frac{1}{\pi}\sum_{n=1}^\infty \frac{(D_\Delta-D_0)(1-\delta)^{2n} \left( r^n + r^{-n}\right) }{n\left\{2D_0 + (D_\Delta-D_0)\left[1-(1-\delta)^{2n}\right]\right\}} \cos(n\theta),\\
\label{Rtwoshells2}
R_2(r,\theta) = \frac{D_\Delta r^2}{4\pi D_0} - \frac{D_\Delta}{8\pi D_0} + \frac{(D_\Delta-D_0)\delta^2(2-\delta)^2}{8\pi D_0} + \frac{1}{\pi} \sum_{n=1}^\infty \frac{(D_\Delta-D_0)\left[(1-\delta)^{2n}+1\right] r^n}{n\left\{2D_0 + (D_\Delta-D_0)\left[1-(1-\delta)^{2n}\right]\right\}}  \cos(n\theta).
\end{gather}
Its value at the center of the escape region ${\bf x_0}$ - $R({\bf x_0}|{\bf x_0}) \equiv R_1(1,0)$ - is then
\begin{equation}
R({\bf x_0}|{\bf x_0})= \frac{1}{8\pi} + \frac{(D_\Delta-D_0)(1-\delta)^4}{8\pi D_0} + \frac{2}{\pi}\sum_{n=1}^\infty \frac{(D_\Delta-D_0)(1-\delta)^{2n}}{n\left\{2D_0 + (D_\Delta-D_0)\left[1-(1-\delta)^{2n}\right]\right\}},
\end{equation}
which is reproduced in Eq.~(\ref{RGreenTL}) of the main text.

In the limit $D_\Delta \rightarrow D_0$, we can remark that the expressions (\ref{Rtwoshells1}-\ref{Rtwoshells2}) become
\begin{equation}
R_1(r,\theta) \simeq R_2(r,\theta) \simeq \frac{r^2}{4\pi} - \frac{1}{8\pi},
\end{equation}
corresponding to the disk formula (\ref{Rdisk}). Moreover, in the limit $\delta \rightarrow 1$, the function $R_1(r,\theta)$ goes also to the disk formula and the function $R_2(r,\theta)$, valid only at $r=0$, is equal to $-1/8\pi$ corresponding to the value of the disk formula at $r=0$. Then, in both limit $D_\Delta \rightarrow D_0$ and $\Delta \rightarrow {\cal R}$, the regular part of the Green’s function tends to the disk formula (\ref{Rdisk}), which implies that the MFPT starting at any position ${\bf x}$ of the domain correctly satisfies the well-known formula (\ref{circularMFPT}).

\section{Simulation of the Langevin equation}
\label{appendixLangevin}

In this appendix we consider the one-dimensional problem on the unit segment $x\in[0,1]$, corresponding to the two-shell geometry with a fully absorbing external boundary ($\varepsilon=2\pi$), where $x$ plays the role of the radial component~$|{\bf x}|/{\cal R}$. The extremity at $x=0$ is reflecting whereas the one at $x=1$ is absorbing. The diffusivity of the particle is denoted as $D_0$ for $x<1-\delta$ and $D_\Delta$ for $x>1-\delta$. The one dimensional MFPT  for a Brownian particle starting at the position $x<1-\delta$ (respectively at $x>1-\delta$) is denoted as $t_1(x)$ (respectively $t_2(x)$). The equations in the bulk are then~\cite{redner_2001_guide}
\begin{gather}
D_0 t_1''(x) = -1, \quad x<1-\delta, \\
D_\Delta t_2''(x) = -1, \quad x>1-\delta.
\end{gather}
The boundary conditions are $t_1'(0) = 0$ and $t_2(1)=0$, and the matching conditions at $x=1-\delta$ are $t_1(1-\delta) = t_2(1-\delta)$ and $D_0 t_1'(1-\delta) = D_\Delta t_2'(1-\delta)$ corresponding respectively to the continuity of the concentration and the flux of diffusive particles across the boundary. The general solution is
\begin{gather}
t_1(x) = \frac{1-x^2}{2 D_0} - \frac{(D_\Delta-D_0)\delta (2-\delta)}{2 D_0 D_\Delta}, \\
t_2(x) = \frac{1-x^2}{2 D_\Delta}.
\end{gather}

The CMFPT expression is given here by
\begin{equation}
\label{CMFPT1d}
t_1(0) = \frac{1}{2 D_0} - \frac{(D_\Delta-D_0)\delta (2-\delta)}{2 D_0 D_\Delta},
\end{equation}
and the GMFPT is then
\begin{equation}
\label{GMFPT1d}
\langle t \rangle = \frac{D_0 \delta [3(1-\delta) + \delta^2] +D_\Delta(1-\delta)^3 }{3 D_0 D_\Delta}.
\end{equation}

We want now to compare these exact one-dimensional solutions to the numerical simulations' ones. We recall here that the general one dimensional Fokker-Planck equation for a diffusive particle with a inhomogeneous diffusivity $D(x)$ writes
\begin{equation}
\frac{\partial p}{\partial t}(x,t) = \frac{\partial}{\partial x} \left[D(x)  \frac{\partial p}{\partial x} (x,t) \right] = -\frac{\partial J}{\partial x}(x,t),
\end{equation}
with a current $J(x,t)$ satisfying the Fick's semi-empirical law. The Langevin equation written with the It\=o convention, corresponding to a full explicit discretization, is
\begin{equation}
\label{langevinIto}
x_{t+dt}-x_t = D'(x_t) dt + \sqrt{2D(x_t)} dB_t 
\end{equation}
where $dB_t$ represents the Wiener process at time $t$ with a variance $dt$. The convective term of the It\=o formalism is then given by $D'(x)$. For our two-shell geometry, with piecewise constant diffusivities, this convective term is thus proportional to the Dirac's distribution.

\begin{figure}[t]
\begin{center}
  \includegraphics[width=8cm]{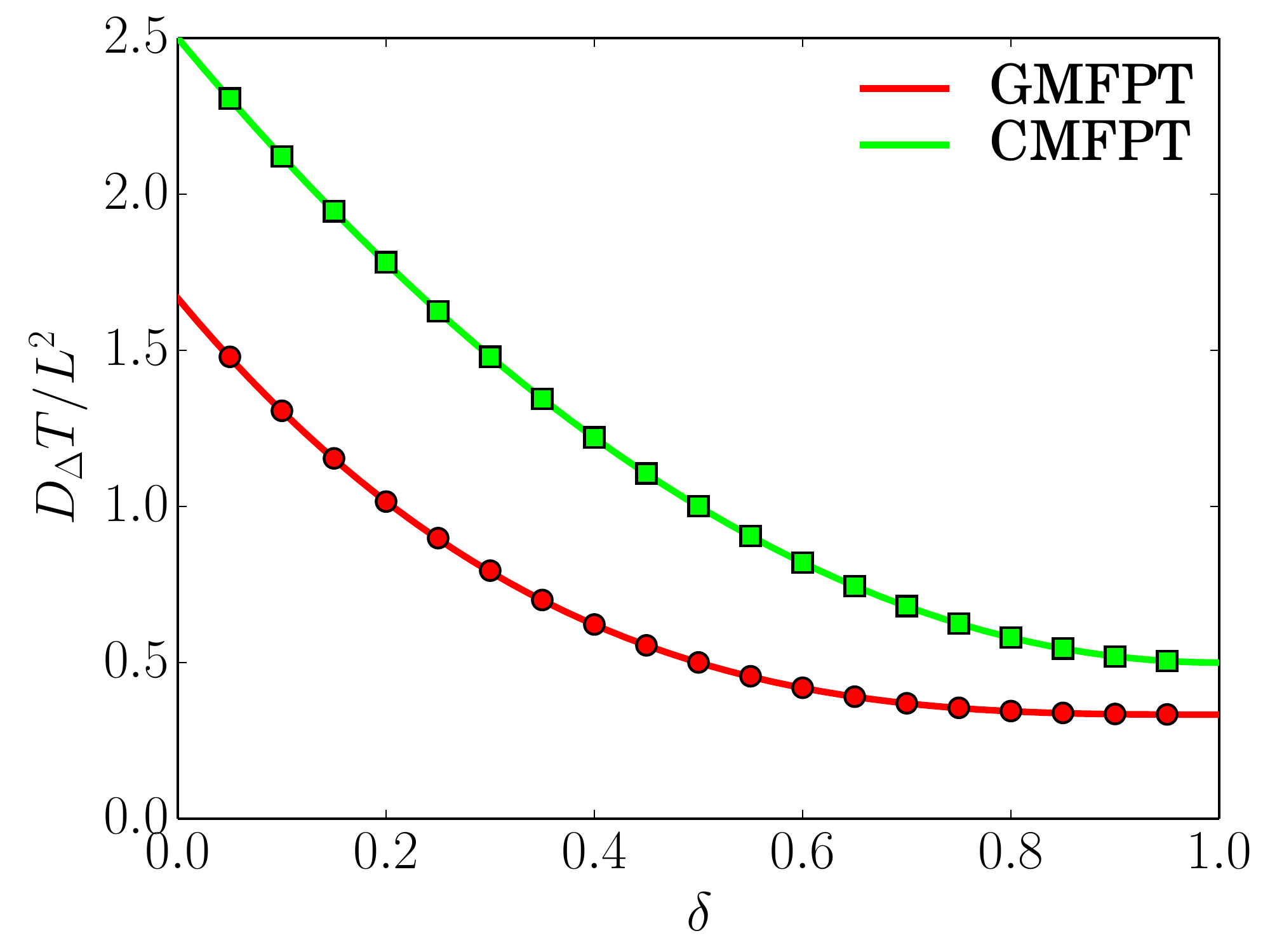}
  \caption{Numerical solution of the dimensionless global mean first passage time (GMFPT) and mean first passage for a particle starting at $x=0$ (CMFPT) plotted respectively with circles and squares as a function of $\delta$ for $D_\Delta/D_0 = 5$. This numerical solution is obtained from the one dimensional numerical simulation whose algorithm is explain in the appendix \ref{appendixLangevin}. The analytical expressions for the CMFPT and the GMFPT given respectively by Eq.~(\ref{CMFPT1d}) and Eq.~(\ref{GMFPT1d}) are plotted with straight lines. The algorithm for discontinuous media shown by~\cite{lejay_2012_simulating, lejay_2013_new} is thus accurate for the one-dimensional problem. \label{fig_langevin1d}}
\end{center}
\end{figure}

To get a realizable numerical simulation for this problem, we follow the algorithm shown by~\cite{lejay_2012_simulating, lejay_2013_new}, satisfying the flux continuity on the inner boundary. When the diffusive particle is exactly at $x=1-\delta$, the probability to go to the left (through a region of diffusivity $D_0$) is
\begin{equation}
\label{pleft}
p_{\rm left} = \frac{\sqrt{D_0}}{ \sqrt{D_0} + \sqrt{D_\Delta}}
\end{equation}
whereas the probability to go to the right (through a region of diffusivity $D_\Delta$) is
\begin{equation}
\label{pright}
p_{\rm right} = \frac{\sqrt{D_\Delta}}{ \sqrt{D_0} + \sqrt{D_\Delta}}.
\end{equation}
If $D_0=D_\Delta$, we get back to a random walker with $p_{\rm left} = p_{\rm right} = 1/2$ equivalent to a diffusive motion with a constant diffusivity. In figure \ref{fig_langevin1d}, we show that this algorithm is compatible with exact expressions of CMFPT and GMFPT given respectively by Eqs.~(\ref{CMFPT1d}) and (\ref{GMFPT1d}) which justify the use of this algorithm for the 2d problem and the numerical data shown in Fig.~\ref{fig_twolayers_NE}.

Remarkably, we can mention that the fully implicit discretization (isothermal convention) writes~\cite{vaccario_2015_firstpassage}
\begin{equation}
x_{t+dt}-x_t = \sqrt{2D(x_{t+dt})} dB_t
\end{equation}
instead of Eq.~(\ref{langevinIto}), with a zero convective term but difficult to integrate numerically.


\begin{thebibliography}{44}
\expandafter\ifx\csname natexlab\endcsname\relax\def\natexlab#1{#1}\fi
\expandafter\ifx\csname bibnamefont\endcsname\relax
  \def\bibnamefont#1{#1}\fi
\expandafter\ifx\csname bibfnamefont\endcsname\relax
  \def\bibfnamefont#1{#1}\fi
\expandafter\ifx\csname citenamefont\endcsname\relax
  \def\citenamefont#1{#1}\fi
\expandafter\ifx\csname url\endcsname\relax
  \def\url#1{\texttt{#1}}\fi
\expandafter\ifx\csname urlprefix\endcsname\relax\def\urlprefix{URL }\fi
\providecommand{\bibinfo}[2]{#2}
\providecommand{\eprint}[2][]{\url{#2}}

\bibitem[{\citenamefont{Schuss et~al.}(2007)\citenamefont{Schuss, Singer, and
  Holcman}}]{schuss_2007_narrow}
\bibinfo{author}{\bibfnamefont{Z.}~\bibnamefont{Schuss}},
  \bibinfo{author}{\bibfnamefont{A.}~\bibnamefont{Singer}}, \bibnamefont{and}
  \bibinfo{author}{\bibfnamefont{D.}~\bibnamefont{Holcman}},
  \bibinfo{journal}{PNAS} \textbf{\bibinfo{volume}{104}},
  \bibinfo{pages}{16098} (\bibinfo{year}{2007}), ISSN \bibinfo{issn}{0027-8424,
  1091-6490}.

\bibitem[{\citenamefont{Bressloff and Newby}(2013)}]{bressloff_2013_stochastic}
\bibinfo{author}{\bibfnamefont{P.~C.} \bibnamefont{Bressloff}}
  \bibnamefont{and} \bibinfo{author}{\bibfnamefont{J.~M.} \bibnamefont{Newby}},
  \bibinfo{journal}{Rev. Mod. Phys.} \textbf{\bibinfo{volume}{85}},
  \bibinfo{pages}{135} (\bibinfo{year}{2013}).

\bibitem[{\citenamefont{Chou and D'Orsogna}(2014)}]{chou_2014_first}
\bibinfo{author}{\bibfnamefont{T.}~\bibnamefont{Chou}} \bibnamefont{and}
  \bibinfo{author}{\bibfnamefont{M.~R.} \bibnamefont{D'Orsogna}},
  \bibinfo{journal}{ArXiv14084518 Cond-Mat Q-Bio}  (\bibinfo{year}{2014}),
  \eprint{1408.4518}.

\bibitem[{\citenamefont{Holcman and Schuss}(2014)}]{holcman_2014_narrow}
\bibinfo{author}{\bibfnamefont{D.}~\bibnamefont{Holcman}} \bibnamefont{and}
  \bibinfo{author}{\bibfnamefont{Z.}~\bibnamefont{Schuss}},
  \bibinfo{journal}{SIAM Rev.} \textbf{\bibinfo{volume}{56}},
  \bibinfo{pages}{213} (\bibinfo{year}{2014}), ISSN \bibinfo{issn}{0036-1445}.

\bibitem[{\citenamefont{{Iyer-Biswas} and
  Zilman}(2015)}]{iyer-biswas_2015_first}
\bibinfo{author}{\bibfnamefont{S.}~\bibnamefont{{Iyer-Biswas}}}
  \bibnamefont{and} \bibinfo{author}{\bibfnamefont{A.}~\bibnamefont{Zilman}},
  \bibinfo{journal}{ArXiv150300291 Cond-Mat Q-Bio}  (\bibinfo{year}{2015}),
  \eprint{1503.00291}.

\bibitem[{\citenamefont{Redner}(2001)}]{redner_2001_guide}
\bibinfo{author}{\bibfnamefont{S.}~\bibnamefont{Redner}},
  \emph{\bibinfo{title}{A {{Guide}} to {{First}}-{{Passage Processes}}}}
  (\bibinfo{publisher}{{Cambridge University Press}}, \bibinfo{year}{2001}),
  ISBN \bibinfo{isbn}{978-0-521-65248-3}.

\bibitem[{\citenamefont{Schwarz and Rieger}(2013)}]{schwarz_2013_efficient}
\bibinfo{author}{\bibfnamefont{K.}~\bibnamefont{Schwarz}} \bibnamefont{and}
  \bibinfo{author}{\bibfnamefont{H.}~\bibnamefont{Rieger}},
  \bibinfo{journal}{Journal of Computational Physics}
  \textbf{\bibinfo{volume}{237}}, \bibinfo{pages}{396} (\bibinfo{year}{2013}),
  ISSN \bibinfo{issn}{0021-9991}.

\bibitem[{\citenamefont{Schwarz
  et~al.}(2016{\natexlab{a}})\citenamefont{Schwarz, Schr\"oder, Qu, Hoth, and
  Rieger}}]{schwarz_2016_optimality}
\bibinfo{author}{\bibfnamefont{K.}~\bibnamefont{Schwarz}},
  \bibinfo{author}{\bibfnamefont{Y.}~\bibnamefont{Schr\"oder}},
  \bibinfo{author}{\bibfnamefont{B.}~\bibnamefont{Qu}},
  \bibinfo{author}{\bibfnamefont{M.}~\bibnamefont{Hoth}}, \bibnamefont{and}
  \bibinfo{author}{\bibfnamefont{H.}~\bibnamefont{Rieger}},
  \bibinfo{journal}{Phys. Rev. Lett.} \textbf{\bibinfo{volume}{117}},
  \bibinfo{pages}{068101} (\bibinfo{year}{2016}{\natexlab{a}}).

\bibitem[{\citenamefont{Schwarz
  et~al.}(2016{\natexlab{b}})\citenamefont{Schwarz, Schr\"oder, and
  Rieger}}]{schwarz_2016_numerical}
\bibinfo{author}{\bibfnamefont{K.}~\bibnamefont{Schwarz}},
  \bibinfo{author}{\bibfnamefont{Y.}~\bibnamefont{Schr\"oder}},
  \bibnamefont{and} \bibinfo{author}{\bibfnamefont{H.}~\bibnamefont{Rieger}},
  \bibinfo{journal}{Phys. Rev. E} \textbf{\bibinfo{volume}{94}},
  \bibinfo{pages}{042133} (\bibinfo{year}{2016}{\natexlab{b}}).

\bibitem[{\citenamefont{Holcman and Schuss}(2004)}]{holcman_2004_escape}
\bibinfo{author}{\bibfnamefont{D.}~\bibnamefont{Holcman}} \bibnamefont{and}
  \bibinfo{author}{\bibfnamefont{Z.}~\bibnamefont{Schuss}},
  \bibinfo{journal}{J. Stat. Phys.} \textbf{\bibinfo{volume}{117}},
  \bibinfo{pages}{975} (\bibinfo{year}{2004}), ISSN \bibinfo{issn}{0022-4715,
  1572-9613}.

\bibitem[{\citenamefont{Kolokolnikov et~al.}(2005)\citenamefont{Kolokolnikov,
  Titcombe, and Ward}}]{kolokolnikov_2005_optimizing}
\bibinfo{author}{\bibfnamefont{T.}~\bibnamefont{Kolokolnikov}},
  \bibinfo{author}{\bibfnamefont{M.~S.} \bibnamefont{Titcombe}},
  \bibnamefont{and} \bibinfo{author}{\bibfnamefont{M.~J.} \bibnamefont{Ward}},
  \bibinfo{journal}{Eur. J. Appl. Math.} \textbf{\bibinfo{volume}{16}},
  \bibinfo{pages}{161} (\bibinfo{year}{2005}), ISSN \bibinfo{issn}{1469-4425,
  0956-7925}.

\bibitem[{\citenamefont{Singer et~al.}(2006{\natexlab{a}})\citenamefont{Singer,
  Schuss, Holcman, and Eisenberg}}]{singer_2006_narrowa}
\bibinfo{author}{\bibfnamefont{A.}~\bibnamefont{Singer}},
  \bibinfo{author}{\bibfnamefont{Z.}~\bibnamefont{Schuss}},
  \bibinfo{author}{\bibfnamefont{D.}~\bibnamefont{Holcman}}, \bibnamefont{and}
  \bibinfo{author}{\bibfnamefont{R.~S.} \bibnamefont{Eisenberg}},
  \bibinfo{journal}{J Stat Phys} \textbf{\bibinfo{volume}{122}},
  \bibinfo{pages}{437} (\bibinfo{year}{2006}{\natexlab{a}}), ISSN
  \bibinfo{issn}{1572-9613}.

\bibitem[{\citenamefont{Singer et~al.}(2006{\natexlab{b}})\citenamefont{Singer,
  Schuss, and Holcman}}]{singer_2006_narrowb}
\bibinfo{author}{\bibfnamefont{A.}~\bibnamefont{Singer}},
  \bibinfo{author}{\bibfnamefont{Z.}~\bibnamefont{Schuss}}, \bibnamefont{and}
  \bibinfo{author}{\bibfnamefont{D.}~\bibnamefont{Holcman}},
  \bibinfo{journal}{J Stat Phys} \textbf{\bibinfo{volume}{122}},
  \bibinfo{pages}{465} (\bibinfo{year}{2006}{\natexlab{b}}), ISSN
  \bibinfo{issn}{1572-9613}.

\bibitem[{\citenamefont{Singer et~al.}(2006{\natexlab{c}})\citenamefont{Singer,
  Schuss, and Holcman}}]{singer_2006_narrowc}
\bibinfo{author}{\bibfnamefont{A.}~\bibnamefont{Singer}},
  \bibinfo{author}{\bibfnamefont{Z.}~\bibnamefont{Schuss}}, \bibnamefont{and}
  \bibinfo{author}{\bibfnamefont{D.}~\bibnamefont{Holcman}},
  \bibinfo{journal}{J Stat Phys} \textbf{\bibinfo{volume}{122}},
  \bibinfo{pages}{491} (\bibinfo{year}{2006}{\natexlab{c}}), ISSN
  \bibinfo{issn}{1572-9613}.

\bibitem[{\citenamefont{Grebenkov}(2016)}]{grebenkov_2016_universal}
\bibinfo{author}{\bibfnamefont{D.~S.} \bibnamefont{Grebenkov}},
  \bibinfo{journal}{Phys. Rev. Lett.} \textbf{\bibinfo{volume}{117}},
  \bibinfo{pages}{260201} (\bibinfo{year}{2016}).

\bibitem[{\citenamefont{Condamin et~al.}(2007)\citenamefont{Condamin,
  B\'enichou, Tejedor, Voituriez, and Klafter}}]{condamin_2007_firstpassage}
\bibinfo{author}{\bibfnamefont{S.}~\bibnamefont{Condamin}},
  \bibinfo{author}{\bibfnamefont{O.}~\bibnamefont{B\'enichou}},
  \bibinfo{author}{\bibfnamefont{V.}~\bibnamefont{Tejedor}},
  \bibinfo{author}{\bibfnamefont{R.}~\bibnamefont{Voituriez}},
  \bibnamefont{and} \bibinfo{author}{\bibfnamefont{J.}~\bibnamefont{Klafter}},
  \bibinfo{journal}{Nature} \textbf{\bibinfo{volume}{450}}, \bibinfo{pages}{77}
  (\bibinfo{year}{2007}).

\bibitem[{\citenamefont{B\'enichou and
  Voituriez}(2008)}]{benichou_2008_narrowescape}
\bibinfo{author}{\bibfnamefont{O.}~\bibnamefont{B\'enichou}} \bibnamefont{and}
  \bibinfo{author}{\bibfnamefont{R.}~\bibnamefont{Voituriez}},
  \bibinfo{journal}{Phys. Rev. Lett.} \textbf{\bibinfo{volume}{100}},
  \bibinfo{pages}{168105} (\bibinfo{year}{2008}).

\bibitem[{\citenamefont{Cheviakov et~al.}(2010)\citenamefont{Cheviakov, Ward,
  and Straube}}]{cheviakov_2010_asymptotic}
\bibinfo{author}{\bibfnamefont{A.}~\bibnamefont{Cheviakov}},
  \bibinfo{author}{\bibfnamefont{M.}~\bibnamefont{Ward}}, \bibnamefont{and}
  \bibinfo{author}{\bibfnamefont{R.}~\bibnamefont{Straube}},
  \bibinfo{journal}{Multiscale Model. Simul.} \textbf{\bibinfo{volume}{8}},
  \bibinfo{pages}{836} (\bibinfo{year}{2010}), ISSN \bibinfo{issn}{1540-3459}.

\bibitem[{\citenamefont{Pillay et~al.}(2010)\citenamefont{Pillay, Ward, Peirce,
  and Kolokolnikov}}]{pillay_2010_asymptotic}
\bibinfo{author}{\bibfnamefont{S.}~\bibnamefont{Pillay}},
  \bibinfo{author}{\bibfnamefont{M.}~\bibnamefont{Ward}},
  \bibinfo{author}{\bibfnamefont{A.}~\bibnamefont{Peirce}}, \bibnamefont{and}
  \bibinfo{author}{\bibfnamefont{T.}~\bibnamefont{Kolokolnikov}},
  \bibinfo{journal}{Multiscale Model. Simul.} \textbf{\bibinfo{volume}{8}},
  \bibinfo{pages}{803} (\bibinfo{year}{2010}), ISSN \bibinfo{issn}{1540-3459}.

\bibitem[{\citenamefont{Chevalier et~al.}(2011)\citenamefont{Chevalier,
  B\'enichou, Meyer, and Voituriez}}]{chevalier_2011_firstpassage}
\bibinfo{author}{\bibfnamefont{C.}~\bibnamefont{Chevalier}},
  \bibinfo{author}{\bibfnamefont{O.}~\bibnamefont{B\'enichou}},
  \bibinfo{author}{\bibfnamefont{B.}~\bibnamefont{Meyer}}, \bibnamefont{and}
  \bibinfo{author}{\bibfnamefont{R.}~\bibnamefont{Voituriez}},
  \bibinfo{journal}{J. Phys. A: Math. Theor.} \textbf{\bibinfo{volume}{44}},
  \bibinfo{pages}{025002} (\bibinfo{year}{2011}), ISSN
  \bibinfo{issn}{1751-8121}.

\bibitem[{\citenamefont{Cheviakov et~al.}(2012)\citenamefont{Cheviakov, Reimer,
  and Ward}}]{cheviakov_2012_mathematical}
\bibinfo{author}{\bibfnamefont{A.~F.} \bibnamefont{Cheviakov}},
  \bibinfo{author}{\bibfnamefont{A.~S.} \bibnamefont{Reimer}},
  \bibnamefont{and} \bibinfo{author}{\bibfnamefont{M.~J.} \bibnamefont{Ward}},
  \bibinfo{journal}{Phys. Rev. E} \textbf{\bibinfo{volume}{85}},
  \bibinfo{pages}{021131} (\bibinfo{year}{2012}).

\bibitem[{\citenamefont{Gomez and Cheviakov}(2015)}]{gomez_2015_asymptotic}
\bibinfo{author}{\bibfnamefont{D.}~\bibnamefont{Gomez}} \bibnamefont{and}
  \bibinfo{author}{\bibfnamefont{A.~F.} \bibnamefont{Cheviakov}},
  \bibinfo{journal}{Phys. Rev. E} \textbf{\bibinfo{volume}{91}},
  \bibinfo{pages}{012137} (\bibinfo{year}{2015}).

\bibitem[{\citenamefont{Hafner and Rieger}(2016)}]{hafner_2016_spatial}
\bibinfo{author}{\bibfnamefont{A.~E.} \bibnamefont{Hafner}} \bibnamefont{and}
  \bibinfo{author}{\bibfnamefont{H.}~\bibnamefont{Rieger}},
  \bibinfo{journal}{Phys. Biol.} \textbf{\bibinfo{volume}{13}},
  \bibinfo{pages}{066003} (\bibinfo{year}{2016}), ISSN
  \bibinfo{issn}{1478-3975}, \eprint{1605.09230}.

\bibitem[{\citenamefont{Hafner and Rieger}(2018)}]{hafner_2018_spatial}
\bibinfo{author}{\bibfnamefont{A.~E.} \bibnamefont{Hafner}} \bibnamefont{and}
  \bibinfo{author}{\bibfnamefont{H.}~\bibnamefont{Rieger}},
  \bibinfo{journal}{Biophysical Journal} \textbf{\bibinfo{volume}{114}},
  \bibinfo{pages}{1420} (\bibinfo{year}{2018}), ISSN \bibinfo{issn}{0006-3495},
  \eprint{1709.05133}.

\bibitem[{\citenamefont{B\'enichou et~al.}(2010)\citenamefont{B\'enichou,
  Grebenkov, Levitz, Loverdo, and Voituriez}}]{benichou_2010_optimal}
\bibinfo{author}{\bibfnamefont{O.}~\bibnamefont{B\'enichou}},
  \bibinfo{author}{\bibfnamefont{D.~S.} \bibnamefont{Grebenkov}},
  \bibinfo{author}{\bibfnamefont{P.~E.} \bibnamefont{Levitz}},
  \bibinfo{author}{\bibfnamefont{C.}~\bibnamefont{Loverdo}}, \bibnamefont{and}
  \bibinfo{author}{\bibfnamefont{R.}~\bibnamefont{Voituriez}},
  \bibinfo{journal}{Phys. Rev. Lett.} \textbf{\bibinfo{volume}{105}},
  \bibinfo{pages}{150606} (\bibinfo{year}{2010}).

\bibitem[{\citenamefont{B\'enichou et~al.}(2011)\citenamefont{B\'enichou,
  Grebenkov, Levitz, Loverdo, and Voituriez}}]{benichou_2011_mean}
\bibinfo{author}{\bibfnamefont{O.}~\bibnamefont{B\'enichou}},
  \bibinfo{author}{\bibfnamefont{D.~S.} \bibnamefont{Grebenkov}},
  \bibinfo{author}{\bibfnamefont{P.~E.} \bibnamefont{Levitz}},
  \bibinfo{author}{\bibfnamefont{C.}~\bibnamefont{Loverdo}}, \bibnamefont{and}
  \bibinfo{author}{\bibfnamefont{R.}~\bibnamefont{Voituriez}},
  \bibinfo{journal}{J. Stat. Phys.} \textbf{\bibinfo{volume}{142}},
  \bibinfo{pages}{657} (\bibinfo{year}{2011}).

\bibitem[{\citenamefont{Calandre et~al.}(2012)\citenamefont{Calandre,
  B\'enichou, Grebenkov, and Voituriez}}]{calandre_2012_interfacial}
\bibinfo{author}{\bibfnamefont{T.}~\bibnamefont{Calandre}},
  \bibinfo{author}{\bibfnamefont{O.}~\bibnamefont{B\'enichou}},
  \bibinfo{author}{\bibfnamefont{D.~S.} \bibnamefont{Grebenkov}},
  \bibnamefont{and}
  \bibinfo{author}{\bibfnamefont{R.}~\bibnamefont{Voituriez}},
  \bibinfo{journal}{Phys. Rev. E} \textbf{\bibinfo{volume}{85}},
  \bibinfo{pages}{051111} (\bibinfo{year}{2012}).

\bibitem[{\citenamefont{Rupprecht
  et~al.}(2012{\natexlab{a}})\citenamefont{Rupprecht, B\'enichou, Grebenkov,
  and Voituriez}}]{rupprecht_2012_kinetics}
\bibinfo{author}{\bibfnamefont{J.-F.} \bibnamefont{Rupprecht}},
  \bibinfo{author}{\bibfnamefont{O.}~\bibnamefont{B\'enichou}},
  \bibinfo{author}{\bibfnamefont{D.~S.} \bibnamefont{Grebenkov}},
  \bibnamefont{and}
  \bibinfo{author}{\bibfnamefont{R.}~\bibnamefont{Voituriez}},
  \bibinfo{journal}{J. Stat. Phys.} \textbf{\bibinfo{volume}{147}},
  \bibinfo{pages}{891} (\bibinfo{year}{2012}{\natexlab{a}}).

\bibitem[{\citenamefont{Rupprecht
  et~al.}(2012{\natexlab{b}})\citenamefont{Rupprecht, B\'enichou, Grebenkov,
  and Voituriez}}]{rupprecht_2012_exact}
\bibinfo{author}{\bibfnamefont{J.-F.} \bibnamefont{Rupprecht}},
  \bibinfo{author}{\bibfnamefont{O.}~\bibnamefont{B\'enichou}},
  \bibinfo{author}{\bibfnamefont{D.~S.} \bibnamefont{Grebenkov}},
  \bibnamefont{and}
  \bibinfo{author}{\bibfnamefont{R.}~\bibnamefont{Voituriez}},
  \bibinfo{journal}{Phys. Rev. E} \textbf{\bibinfo{volume}{86}},
  \bibinfo{pages}{041135} (\bibinfo{year}{2012}{\natexlab{b}}).

\bibitem[{\citenamefont{LaBolle et~al.}(2000)\citenamefont{LaBolle, Quastel,
  Fogg, and Gravner}}]{labolle_2000_diffusion}
\bibinfo{author}{\bibfnamefont{E.~M.} \bibnamefont{LaBolle}},
  \bibinfo{author}{\bibfnamefont{J.}~\bibnamefont{Quastel}},
  \bibinfo{author}{\bibfnamefont{G.~E.} \bibnamefont{Fogg}}, \bibnamefont{and}
  \bibinfo{author}{\bibfnamefont{J.}~\bibnamefont{Gravner}},
  \bibinfo{journal}{Water Resour. Res.} \textbf{\bibinfo{volume}{36}},
  \bibinfo{pages}{651} (\bibinfo{year}{2000}), ISSN \bibinfo{issn}{1944-7973}.

\bibitem[{\citenamefont{{Alvarez-Ramirez}
  et~al.}(2014{\natexlab{a}})\citenamefont{{Alvarez-Ramirez}, Dagdug, and
  Meraz}}]{alvarez-ramirez_2014_asymmetric}
\bibinfo{author}{\bibfnamefont{J.}~\bibnamefont{{Alvarez-Ramirez}}},
  \bibinfo{author}{\bibfnamefont{L.}~\bibnamefont{Dagdug}}, \bibnamefont{and}
  \bibinfo{author}{\bibfnamefont{M.}~\bibnamefont{Meraz}},
  \bibinfo{journal}{Physica A: Statistical Mechanics and its Applications}
  \textbf{\bibinfo{volume}{395}}, \bibinfo{pages}{193}
  (\bibinfo{year}{2014}{\natexlab{a}}), ISSN \bibinfo{issn}{0378-4371}.

\bibitem[{\citenamefont{{Alvarez-Ramirez}
  et~al.}(2014{\natexlab{b}})\citenamefont{{Alvarez-Ramirez}, Dagdug, Inzunza,
  and Rodriguez}}]{alvarez-ramirez_2014_asymmetrical}
\bibinfo{author}{\bibfnamefont{J.}~\bibnamefont{{Alvarez-Ramirez}}},
  \bibinfo{author}{\bibfnamefont{L.}~\bibnamefont{Dagdug}},
  \bibinfo{author}{\bibfnamefont{L.}~\bibnamefont{Inzunza}}, \bibnamefont{and}
  \bibinfo{author}{\bibfnamefont{E.}~\bibnamefont{Rodriguez}},
  \bibinfo{journal}{Physica A: Statistical Mechanics and its Applications}
  \textbf{\bibinfo{volume}{407}}, \bibinfo{pages}{24}
  (\bibinfo{year}{2014}{\natexlab{b}}), ISSN \bibinfo{issn}{0378-4371}.

\bibitem[{\citenamefont{Lejay and Martinez}(2006)}]{lejay_2006_scheme}
\bibinfo{author}{\bibfnamefont{A.}~\bibnamefont{Lejay}} \bibnamefont{and}
  \bibinfo{author}{\bibfnamefont{M.}~\bibnamefont{Martinez}},
  \bibinfo{journal}{Ann. Appl. Probab.} \textbf{\bibinfo{volume}{16}},
  \bibinfo{pages}{107} (\bibinfo{year}{2006}), ISSN \bibinfo{issn}{1050-5164,
  2168-8737}.

\bibitem[{\citenamefont{Lejay and Pichot}(2012)}]{lejay_2012_simulating}
\bibinfo{author}{\bibfnamefont{A.}~\bibnamefont{Lejay}} \bibnamefont{and}
  \bibinfo{author}{\bibfnamefont{G.}~\bibnamefont{Pichot}},
  \bibinfo{journal}{Journal of Computational Physics}
  \textbf{\bibinfo{volume}{231}}, \bibinfo{pages}{7299} (\bibinfo{year}{2012}),
  ISSN \bibinfo{issn}{0021-9991}.

\bibitem[{\citenamefont{Lejay and Maire}(2013)}]{lejay_2013_new}
\bibinfo{author}{\bibfnamefont{A.}~\bibnamefont{Lejay}} \bibnamefont{and}
  \bibinfo{author}{\bibfnamefont{S.}~\bibnamefont{Maire}},
  \bibinfo{journal}{Journal of Computational and Applied Mathematics}
  \textbf{\bibinfo{volume}{245}}, \bibinfo{pages}{97} (\bibinfo{year}{2013}),
  ISSN \bibinfo{issn}{0377-0427}.

\bibitem[{\citenamefont{Vaccario et~al.}(2015)\citenamefont{Vaccario, Antoine,
  and Talbot}}]{vaccario_2015_firstpassage}
\bibinfo{author}{\bibfnamefont{G.}~\bibnamefont{Vaccario}},
  \bibinfo{author}{\bibfnamefont{C.}~\bibnamefont{Antoine}}, \bibnamefont{and}
  \bibinfo{author}{\bibfnamefont{J.}~\bibnamefont{Talbot}},
  \bibinfo{journal}{Phys. Rev. Lett.} \textbf{\bibinfo{volume}{115}},
  \bibinfo{pages}{240601} (\bibinfo{year}{2015}).

\bibitem[{\citenamefont{Hecht}(2013)}]{hecht_2013_new}
\bibinfo{author}{\bibfnamefont{F.}~\bibnamefont{Hecht}}, \bibinfo{journal}{J.
  Numer. Math.} \textbf{\bibinfo{volume}{20}}, \bibinfo{pages}{251}
  (\bibinfo{year}{2013}), ISSN \bibinfo{issn}{1569-3953}.

\bibitem[{\citenamefont{Caginalp and Chen}(2012)}]{caginalp_2012_analytical}
\bibinfo{author}{\bibfnamefont{C.}~\bibnamefont{Caginalp}} \bibnamefont{and}
  \bibinfo{author}{\bibfnamefont{X.}~\bibnamefont{Chen}},
  \bibinfo{journal}{Arch Rational Mech Anal} \textbf{\bibinfo{volume}{203}},
  \bibinfo{pages}{329} (\bibinfo{year}{2012}), ISSN \bibinfo{issn}{1432-0673}.

\bibitem[{\citenamefont{Rupprecht et~al.}(2015)\citenamefont{Rupprecht,
  B\'enichou, Grebenkov, and Voituriez}}]{rupprecht_2015_exit}
\bibinfo{author}{\bibfnamefont{J.-F.} \bibnamefont{Rupprecht}},
  \bibinfo{author}{\bibfnamefont{O.}~\bibnamefont{B\'enichou}},
  \bibinfo{author}{\bibfnamefont{D.~S.} \bibnamefont{Grebenkov}},
  \bibnamefont{and}
  \bibinfo{author}{\bibfnamefont{R.}~\bibnamefont{Voituriez}},
  \bibinfo{journal}{J Stat Phys} \textbf{\bibinfo{volume}{158}},
  \bibinfo{pages}{192} (\bibinfo{year}{2015}), ISSN \bibinfo{issn}{1572-9613}.

\bibitem[{\citenamefont{Ryabov et~al.}(2015)\citenamefont{Ryabov, Berestneva,
  and Holubec}}]{ryabov_2015_brownian}
\bibinfo{author}{\bibfnamefont{A.}~\bibnamefont{Ryabov}},
  \bibinfo{author}{\bibfnamefont{E.}~\bibnamefont{Berestneva}},
  \bibnamefont{and} \bibinfo{author}{\bibfnamefont{V.}~\bibnamefont{Holubec}},
  \bibinfo{journal}{J. Chem. Phys.} \textbf{\bibinfo{volume}{143}},
  \bibinfo{pages}{114117} (\bibinfo{year}{2015}), ISSN
  \bibinfo{issn}{0021-9606}.

\bibitem[{\citenamefont{Grebenkov et~al.}(2017)\citenamefont{Grebenkov,
  Metzler, and Oshanin}}]{grebenkov_2017_effects}
\bibinfo{author}{\bibfnamefont{D.~S.} \bibnamefont{Grebenkov}},
  \bibinfo{author}{\bibfnamefont{R.}~\bibnamefont{Metzler}}, \bibnamefont{and}
  \bibinfo{author}{\bibfnamefont{G.}~\bibnamefont{Oshanin}},
  \bibinfo{journal}{New J. Phys.} \textbf{\bibinfo{volume}{19}},
  \bibinfo{pages}{103025} (\bibinfo{year}{2017}), ISSN
  \bibinfo{issn}{1367-2630}.

\bibitem[{\citenamefont{Grebenkov and
  Oshanin}(2017)}]{grebenkov_2017_diffusive}
\bibinfo{author}{\bibfnamefont{D.~S.} \bibnamefont{Grebenkov}}
  \bibnamefont{and} \bibinfo{author}{\bibfnamefont{G.}~\bibnamefont{Oshanin}},
  \bibinfo{journal}{Phys. Chem. Chem. Phys.} \textbf{\bibinfo{volume}{19}},
  \bibinfo{pages}{2723} (\bibinfo{year}{2017}), ISSN \bibinfo{issn}{1463-9084}.

\bibitem[{\citenamefont{Majumdar et~al.}(2015)\citenamefont{Majumdar,
  Sabhapandit, and Schehr}}]{majumdar_2015_random}
\bibinfo{author}{\bibfnamefont{S.~N.} \bibnamefont{Majumdar}},
  \bibinfo{author}{\bibfnamefont{S.}~\bibnamefont{Sabhapandit}},
  \bibnamefont{and} \bibinfo{author}{\bibfnamefont{G.}~\bibnamefont{Schehr}},
  \bibinfo{journal}{Phys. Rev. E} \textbf{\bibinfo{volume}{92}},
  \bibinfo{pages}{052126} (\bibinfo{year}{2015}).

\bibitem[{\citenamefont{Ward and Keller}(1993)}]{ward_1993_strong}
\bibinfo{author}{\bibfnamefont{M.~J.} \bibnamefont{Ward}} \bibnamefont{and}
  \bibinfo{author}{\bibfnamefont{J.~B.} \bibnamefont{Keller}},
  \bibinfo{journal}{SIAM J. Appl. Math.} \textbf{\bibinfo{volume}{53}},
  \bibinfo{pages}{770} (\bibinfo{year}{1993}).

\end{thebibliography}
\end{document}